\documentclass[twocolumn,longauth, traditabstract]{aa}

\usepackage[nonamebreak]{natbib}
\usepackage[stable]{footmisc}
\usepackage{amsmath}
\usepackage{txfonts}
\usepackage{placeins}
\bibpunct{(}{)}{;}{a}{}{,}
\usepackage{graphicx}
\usepackage{epstopdf}
\usepackage{ifthen} 
\usepackage{overpic}
\usepackage{rotating}
\usepackage{subfigure}
\usepackage{epsfig}
\usepackage{multirow}
\usepackage{array}
\usepackage{color}
\usepackage[usernames,dvipsnames]{xcolor}
\usepackage{psfrag}
\usepackage[T1]{fontenc}
\usepackage{url}
\usepackage{breakurl}
\def\UrlBreaks{\do\ }

\usepackage[colorlinks, citecolor=blue]{hyperref}

\newcolumntype{L}[1]{>{\raggedright\let\newline\\\arraybackslash\hspace{0pt}}m{#1}}
\newcolumntype{C}[1]{>{\centering\let\newline\\\arraybackslash\hspace{0pt}}m{#1}}
\newcolumntype{R}[1]{>{\raggedleft\let\newline\\\arraybackslash\hspace{0pt}}m{#1}}

\def\setsymbol#1#2{\expandafter\def\csname #1\endcsname{#2}}
\def\getsymbol#1{\csname #1\endcsname}

\def\Planck{\textit{Planck}}

\newbox\tablebox    \newdimen\tablewidth
\def\leaderfil{\leaders\hbox to 5pt{\hss.\hss}\hfil}
\def\endPlancktablewide{\tablewidth=\textwidth 
    $$\hss\copy\tablebox\hss$$
    \vskip-\lastskip\vskip -2pt}
\def\tablenote#1 #2\par{\begingroup \parindent=0.8em
    \abovedisplayshortskip=0pt\belowdisplayshortskip=0pt
    \noindent
    $$\hss\vbox{\hsize\tablewidth \hangindent=\parindent \hangafter=1 \noindent
    \hbox to \parindent{$^#1$\hss}\strut#2\strut\par}\hss$$
    \endgroup}
\def\doubleline{\vskip 3pt\hrule \vskip 1.5pt \hrule \vskip 5pt}

\def\L2{\ifmmode L_2\else $L_2$\fi}

\def\DeltaT{\ifmmode \Delta T\else $\Delta T$\fi}
\def\deltat{\ifmmode \Delta t\else $\Delta t$\fi}
\def\fknee{\ifmmode f_{\rm knee}\else $f_{\rm knee}$\fi}
\def\Fmax{\ifmmode F_{\rm max}\else $F_{\rm max}$\fi}
\def\solar{\ifmmode{\rm M}_{\mathord\odot}\else${\rm M}_{\mathord\odot}$\fi}
\def\Msolar{\ifmmode{\rm M}_{\mathord\odot}\else${\rm M}_{\mathord\odot}$\fi}
\def\Lsolar{\ifmmode{\rm L}_{\mathord\odot}\else${\rm L}_{\mathord\odot}$\fi}
\def\inv{\ifmmode^{-1}\else$^{-1}$\fi}
\def\mo{\ifmmode^{-1}\else$^{-1}$\fi}
\def\sup#1{\ifmmode ^{\rm #1}\else $^{\rm #1}$\fi}
\def\expo#1{\ifmmode \times 10^{#1}\else $\times 10^{#1}$\fi}
\def\,{\thinspace}
\def\lsim{\mathrel{\raise .4ex\hbox{\rlap{$<$}\lower 1.2ex\hbox{$\sim$}}}}
\def\gsim{\mathrel{\raise .4ex\hbox{\rlap{$>$}\lower 1.2ex\hbox{$\sim$}}}}

\def\simprop{\mathrel{\raise .4ex\hbox{\rlap{$\propto$}\lower 1.2ex\hbox{$\sim$}}}}
\def\deg{\ifmmode^\circ\else$^\circ$\fi}
\def\pdeg{\ifmmode $\setbox0=\hbox{$^{\circ}$}\rlap{\hskip.11\wd0 .}$^{\circ}
          \else \setbox0=\hbox{$^{\circ}$}\rlap{\hskip.11\wd0 .}$^{\circ}$\fi}
\def\arcs{\ifmmode {^{\scriptstyle\prime\prime}}
          \else $^{\scriptstyle\prime\prime}$\fi}
\def\arcm{\ifmmode {^{\scriptstyle\prime}}
          \else $^{\scriptstyle\prime}$\fi}
\newdimen\sa  \newdimen\sb
\def\parcs{\sa=.07em \sb=.03em
     \ifmmode \hbox{\rlap{.}}^{\scriptstyle\prime\kern -\sb\prime}\hbox{\kern -\sa}
     \else \rlap{.}$^{\scriptstyle\prime\kern -\sb\prime}$\kern -\sa\fi}
\def\parcm{\sa=.08em \sb=.03em
     \ifmmode \hbox{\rlap{.}\kern\sa}^{\scriptstyle\prime}\hbox{\kern-\sb}
     \else \rlap{.}\kern\sa$^{\scriptstyle\prime}$\kern-\sb\fi}
\def\ra[#1 #2 #3.#4]{#1\sup{h}#2\sup{m}#3\sup{s}\llap.#4}
\def\dec[#1 #2 #3.#4]{#1\deg#2\arcm#3\arcs\llap.#4}
\def\deco[#1 #2 #3]{#1\deg#2\arcm#3\arcs}
\def\rra[#1 #2]{#1\sup{h}#2\sup{m}}

\def\dots{\relax\ifmmode \ldots\else $\ldots$\fi}
\def\WHzsr{\ifmmode $W\,Hz\mo\,sr\mo$\else W\,Hz\mo\,sr\mo\fi}
\def\mHz{\ifmmode $\,mHz$\else \,mHz\fi}
\def\GHz{\ifmmode $\,GHz$\else \,GHz\fi}
\def\mKs{\ifmmode $\,mK\,s$^{1/2}\else \,mK\,s$^{1/2}$\fi}
\def\muKs{\ifmmode \,\mu$K\,s$^{1/2}\else \,$\mu$K\,s$^{1/2}$\fi}
\def\muKRJs{\ifmmode \,\mu$K$_{\rm RJ}$\,s$^{1/2}\else \,$\mu$K$_{\rm RJ}$\,s$^{1/2}$\fi}
\def\muKHz{\ifmmode \,\mu$K\,Hz$^{-1/2}\else \,$\mu$K\,Hz$^{-1/2}$\fi}
\def\MJysr{\ifmmode \,$MJy\,sr\mo$\else \,MJy\,sr\mo\fi}
\def\MJysrmK{\ifmmode \,$MJy\,sr\mo$\,mK$_{\rm CMB}\mo\else \,MJy\,sr\mo\,mK$_{\rm CMB}\mo$\fi}
\def\microns{\ifmmode \,\mu$m$\else \,$\mu$m\fi}

\def\muK{\ifmmode \,\mu$K$\else \,$\mu$\hbox{K}\fi}
\def\microK{\ifmmode \,\mu$K$\else \,$\mu$\hbox{K}\fi}
\def\muW{\ifmmode \,\mu$W$\else \,$\mu$\hbox{W}\fi}
\def\kms{\ifmmode $\,km\,s$^{-1}\else \,km\,s$^{-1}$\fi}
\def\kmsMpc{\ifmmode $\,\kms\,Mpc\mo$\else \,\kms\,Mpc\mo\fi}
\providecommand{\sorthelp}[1]{}

\newcommand{\hccor}[1]{}
\newcommand{\hccom}[1]{}

\newcommand{\hatn}{\vec{\hat{n\,}}}

\begin{document}

\topmargin=-1cm
\oddsidemargin=-1cm
\evensidemargin=-1cm
\textwidth=17cm
\textheight=25cm
\raggedbottom
\sloppy

\definecolor{Blue}{rgb}{0.,0.,1.}
\definecolor{LightSkyBlue}{rgb}{0.691,0.827,1.}
\definecolor{Red}{rgb}{1.,0.,0.}
\definecolor{Green}{rgb}{0.,1.,0.}
\definecolor{Purple}{rgb}{0.5, 0., 0.5}
\definecolor{Try}{rgb}{0.15,0.,1}
\definecolor{Black}{rgb}{0., 0., 0.}

\title{\textit{Planck} intermediate results. L.
Evidence for spatial\\ variation of the polarized thermal dust spectral energy\\
distribution and implications for CMB $B$-mode analysis}

\author{\small
Planck Collaboration: N.~Aghanim\inst{51}
\and
M.~Ashdown\inst{61, 6}
\and
J.~Aumont\inst{51}$^{*}$
\and
C.~Baccigalupi\inst{74}
\and
M.~Ballardini\inst{25, 41, 44}
\and
A.~J.~Banday\inst{82, 9}
\and
R.~B.~Barreiro\inst{56}
\and
N.~Bartolo\inst{24, 57}
\and
S.~Basak\inst{74}
\and
K.~Benabed\inst{52, 81}
% ...
}
\institute{\small
APC, AstroParticule et Cosmologie, Universit\'{e} Paris Diderot, CNRS/IN2P3, CEA/lrfu, Observatoire de Paris, Sorbonne Paris Cit\'{e}, 10, rue Alice Domon et L\'{e}onie Duquet, 75205 Paris Cedex 13, France\goodbreak
\and
Aalto University Mets\"{a}hovi Radio Observatory and Dept of Radio Science and Engineering, P.O. Box 13000, FI-00076 AALTO, Finland\goodbreak
\and
African Institute for Mathematical Sciences, 6-8 Melrose Road, Muizenberg, Cape Town, South Africa\goodbreak
\and
Agenzia Spaziale Italiana Science Data Center, Via del Politecnico snc, 00133, Roma, Italy\goodbreak
\and
Aix Marseille Universit\'{e}, CNRS, LAM (Laboratoire d'Astrophysique de Marseille) UMR 7326, 13388, Marseille, France\goodbreak
\and
Astrophysics Group, Cavendish Laboratory, University of Cambridge, J J Thomson Avenue, Cambridge CB3 0HE, U.K.\goodbreak
\and
Astrophysics \& Cosmology Research Unit, School of Mathematics, Statistics \& Computer Science, University of KwaZulu-Natal, Westville Campus, Private Bag X54001, Durban 4000, South Africa\goodbreak
\and
CITA, University of Toronto, 60 St. George St., Toronto, ON M5S 3H8, Canada\goodbreak
\and
CNRS, IRAP, 9 Av. colonel Roche, BP 44346, F-31028 Toulouse cedex 4, France\goodbreak
\and
California Institute of Technology, Pasadena, California, U.S.A.\goodbreak
\and
Computational Cosmology Center, Lawrence Berkeley National Laboratory, Berkeley, California, U.S.A.\goodbreak
\and
DTU Space, National Space Institute, Technical University of Denmark, Elektrovej 327, DK-2800 Kgs. Lyngby, Denmark\goodbreak
\and
D\'{e}partement de Physique Th\'{e}orique, Universit\'{e} de Gen\`{e}ve, 24, Quai E. Ansermet,1211 Gen\`{e}ve 4, Switzerland\goodbreak
\and
Departamento de Astrof\'{i}sica, Universidad de La Laguna (ULL), E-38206 La Laguna, Tenerife, Spain\goodbreak
\and
Departamento de F\'{\i}sica, Universidad de Oviedo, Avda. Calvo Sotelo s/n, Oviedo, Spain\goodbreak
\and
Department of Astrophysics/IMAPP, Radboud University Nijmegen, P.O. Box 9010, 6500 GL Nijmegen, The Netherlands\goodbreak
\and
Department of Physics \& Astronomy, University of British Columbia, 6224 Agricultural Road, Vancouver, British Columbia, Canada\goodbreak
\and
Department of Physics and Astronomy, Dana and David Dornsife College of Letter, Arts and Sciences, University of Southern California, Los Angeles, CA 90089, U.S.A.\goodbreak
\and
Department of Physics and Astronomy, University College London, London WC1E 6BT, U.K.\goodbreak
\and
Department of Physics, Gustaf H\"{a}llstr\"{o}min katu 2a, University of Helsinki, Helsinki, Finland\goodbreak
\and
Department of Physics, Princeton University, Princeton, New Jersey, U.S.A.\goodbreak
\and
Department of Physics, University of California, Santa Barbara, California, U.S.A.\goodbreak
\and
Department of Physics, University of Illinois at Urbana-Champaign, 1110 West Green Street, Urbana, Illinois, U.S.A.\goodbreak
\and
Dipartimento di Fisica e Astronomia G. Galilei, Universit\`{a} degli Studi di Padova, via Marzolo 8, 35131 Padova, Italy\goodbreak
\and
Dipartimento di Fisica e Astronomia, Alma Mater Studiorum, Universit\`{a} degli Studi di Bologna, Viale Berti Pichat 6/2, I-40127, Bologna, Italy\goodbreak
\and
Dipartimento di Fisica e Scienze della Terra, Universit\`{a} di Ferrara, Via Saragat 1, 44122 Ferrara, Italy\goodbreak
\and
Dipartimento di Fisica, Universit\`{a} La Sapienza, P. le A. Moro 2, Roma, Italy\goodbreak
\and
Dipartimento di Fisica, Universit\`{a} degli Studi di Milano, Via Celoria, 16, Milano, Italy\goodbreak
\and
Dipartimento di Fisica, Universit\`{a} degli Studi di Trieste, via A. Valerio 2, Trieste, Italy\goodbreak
\and
Dipartimento di Matematica, Universit\`{a} di Roma Tor Vergata, Via della Ricerca Scientifica, 1, Roma, Italy\goodbreak
\and
Discovery Center, Niels Bohr Institute, Copenhagen University, Blegdamsvej 17, Copenhagen, Denmark\goodbreak
\and
European Space Agency, ESAC, Planck Science Office, Camino bajo del Castillo, s/n, Urbanizaci\'{o}n Villafranca del Castillo, Villanueva de la Ca\~{n}ada, Madrid, Spain\goodbreak
\and
European Space Agency, ESTEC, Keplerlaan 1, 2201 AZ Noordwijk, The Netherlands\goodbreak
\and
Gran Sasso Science Institute, INFN, viale F. Crispi 7, 67100 L'Aquila, Italy\goodbreak
\and
HGSFP and University of Heidelberg, Theoretical Physics Department, Philosophenweg 16, 69120, Heidelberg, Germany\goodbreak
\and
Haverford College Astronomy Department, 370 Lancaster Avenue, Haverford, Pennsylvania, U.S.A.\goodbreak
\and
Helsinki Institute of Physics, Gustaf H\"{a}llstr\"{o}min katu 2, University of Helsinki, Helsinki, Finland\goodbreak
\and
INAF - Osservatorio Astronomico di Padova, Vicolo dell'Osservatorio 5, Padova, Italy\goodbreak
\and
INAF - Osservatorio Astronomico di Roma, via di Frascati 33, Monte Porzio Catone, Italy\goodbreak
\and
INAF - Osservatorio Astronomico di Trieste, Via G.B. Tiepolo 11, Trieste, Italy\goodbreak
\and
INAF/IASF Bologna, Via Gobetti 101, Bologna, Italy\goodbreak
\and
INAF/IASF Milano, Via E. Bassini 15, Milano, Italy\goodbreak
\and
INFN - CNAF, viale Berti Pichat 6/2, 40127 Bologna, Italy\goodbreak
\and
INFN, Sezione di Bologna, viale Berti Pichat 6/2, 40127 Bologna, Italy\goodbreak
\and
INFN, Sezione di Ferrara, Via Saragat 1, 44122 Ferrara, Italy\goodbreak
\and
INFN, Sezione di Roma 1, Universit\`{a} di Roma Sapienza, Piazzale Aldo Moro 2, 00185, Roma, Italy\goodbreak
\and
INFN, Sezione di Roma 2, Universit\`{a} di Roma Tor Vergata, Via della Ricerca Scientifica, 1, Roma, Italy\goodbreak
\and
INFN/National Institute for Nuclear Physics, Via Valerio 2, I-34127 Trieste, Italy\goodbreak
\and
IUCAA, Post Bag 4, Ganeshkhind, Pune University Campus, Pune 411 007, India\goodbreak
\and
Imperial College London, Astrophysics group, Blackett Laboratory, Prince Consort Road, London, SW7 2AZ, U.K.\goodbreak
\and
Institut d'Astrophysique Spatiale, CNRS, Univ. Paris-Sud, Universit\'{e} Paris-Saclay, B\^{a}t. 121, 91405 Orsay cedex, France\goodbreak
\and
Institut d'Astrophysique de Paris, CNRS (UMR7095), 98 bis Boulevard Arago, F-75014, Paris, France\goodbreak
\and
Institute of Astronomy, University of Cambridge, Madingley Road, Cambridge CB3 0HA, U.K.\goodbreak
\and
Institute of Theoretical Astrophysics, University of Oslo, Blindern, Oslo, Norway\goodbreak
\and
Instituto de Astrof\'{\i}sica de Canarias, C/V\'{\i}a L\'{a}ctea s/n, La Laguna, Tenerife, Spain\goodbreak
\and
Instituto de F\'{\i}sica de Cantabria (CSIC-Universidad de Cantabria), Avda. de los Castros s/n, Santander, Spain\goodbreak
\and
Istituto Nazionale di Fisica Nucleare, Sezione di Padova, via Marzolo 8, I-35131 Padova, Italy\goodbreak
\and
Jet Propulsion Laboratory, California Institute of Technology, 4800 Oak Grove Drive, Pasadena, California, U.S.A.\goodbreak
\and
Jodrell Bank Centre for Astrophysics, Alan Turing Building, School of Physics and Astronomy, The University of Manchester, Oxford Road, Manchester, M13 9PL, U.K.\goodbreak
\and
Kavli Institute for Cosmological Physics, University of Chicago, Chicago, IL 60637, USA\goodbreak
\and
Kavli Institute for Cosmology Cambridge, Madingley Road, Cambridge, CB3 0HA, U.K.\goodbreak
\and
LAL, Universit\'{e} Paris-Sud, CNRS/IN2P3, Orsay, France\goodbreak
\and
LERMA, CNRS, Observatoire de Paris, 61 Avenue de l'Observatoire, Paris, France\goodbreak
\and
Laboratoire AIM, IRFU/Service d'Astrophysique - CEA/DSM - CNRS - Universit\'{e} Paris Diderot, B\^{a}t. 709, CEA-Saclay, F-91191 Gif-sur-Yvette Cedex, France\goodbreak
\and
Laboratoire Traitement et Communication de l'Information, CNRS (UMR 5141) and T\'{e}l\'{e}com ParisTech, 46 rue Barrault F-75634 Paris Cedex 13, France\goodbreak
\and
Laboratoire de Physique Subatomique et Cosmologie, Universit\'{e} Grenoble-Alpes, CNRS/IN2P3, 53, rue des Martyrs, 38026 Grenoble Cedex, France\goodbreak
\and
Laboratoire de Physique Th\'{e}orique, Universit\'{e} Paris-Sud 11 \& CNRS, B\^{a}timent 210, 91405 Orsay, France\goodbreak
\and
Lawrence Berkeley National Laboratory, Berkeley, California, U.S.A.\goodbreak
\and
Max-Planck-Institut f\"{u}r Astrophysik, Karl-Schwarzschild-Str. 1, 85741 Garching, Germany\goodbreak
\and
Mullard Space Science Laboratory, University College London, Surrey RH5 6NT, U.K.\goodbreak
\and
Nicolaus Copernicus Astronomical Center, Bartycka 18, 00-716 Warsaw, Poland\goodbreak
\and
Niels Bohr Institute, Copenhagen University, Blegdamsvej 17, Copenhagen, Denmark\goodbreak
\and
Nordita (Nordic Institute for Theoretical Physics), Roslagstullsbacken 23, SE-106 91 Stockholm, Sweden\goodbreak
\and
SISSA, Astrophysics Sector, via Bonomea 265, 34136, Trieste, Italy\goodbreak
\and
School of Physics and Astronomy, University of Nottingham, Nottingham NG7 2RD, U.K.\goodbreak
\and
Simon Fraser University, Department of Physics, 8888 University Drive, Burnaby BC, Canada\goodbreak
\and
Sorbonne Universit\'{e}-UPMC, UMR7095, Institut d'Astrophysique de Paris, 98 bis Boulevard Arago, F-75014, Paris, France\goodbreak
\and
Space Sciences Laboratory, University of California, Berkeley, California, U.S.A.\goodbreak
\and
Sub-Department of Astrophysics, University of Oxford, Keble Road, Oxford OX1 3RH, U.K.\goodbreak
\and
The Oskar Klein Centre for Cosmoparticle Physics, Department of Physics,Stockholm University, AlbaNova, SE-106 91 Stockholm, Sweden\goodbreak
\and
UPMC Univ Paris 06, UMR7095, 98 bis Boulevard Arago, F-75014, Paris, France\goodbreak
\and
Universit\'{e} de Toulouse, UPS-OMP, IRAP, F-31028 Toulouse cedex 4, France\goodbreak
% ...
}

\abstract{
The characterization of the Galactic foregrounds has been shown to 
be the main obstacle in the challenging quest to detect primordial $B$-modes
in the polarized microwave sky. 
We make use of the \Planck-HFI 2015 data release at high frequencies to place
new constraints on the properties of the polarized thermal dust emission at high
Galactic latitudes.  Here, we specifically study the spatial variability of the
dust polarized spectral energy distribution (SED), and its potential impact 
on the determination of the tensor-to-scalar ratio, $r$. We use the correlation 
ratio of the $C_\ell^{BB}$ angular power spectra between the 217- and 353-GHz 
channels as a tracer of these potential variations, computed on different high
Galactic latitude regions, ranging from 80\,\% to 20\,\% of the sky. 
The new insight from \Planck\ data is a departure of the correlation ratio
from unity that cannot be attributed to a spurious decorrelation 
due to the cosmic microwave background, instrumental noise, or instrumental
systematics. The effect is marginally detected on each region, but the
statistical combination of all the regions gives more than 99\,\% confidence
for this variation in polarized dust properties.
In addition, we show that the decorrelation increases when there is a decrease
in the mean column density of the region of the sky being considered, and we
propose a simple power-law empirical model for this dependence, which matches
what is seen in the \Planck\ data. We explore the effect that this measured
decorrelation has on simulations of the BICEP2-{\it Keck} Array/\Planck\
analysis and show that the 2015 constraints from those data still allow a
decorrelation between the dust at 150 and 353\,GHz of the order of the one we
measure. Finally, using simplified models, we show that either spatial
variation of the dust SED or of the dust polarization angle could produce
decorrelations between 217- and 353-GHz data similar to those we observe in
the data.
}

\authorrunning{Planck Collaboration}
\titlerunning{Spatial variation of the polarized dust spectral energy distribution}
\maketitle

\section{Introduction}

Since the combined BICEP2/{\it Keck Array} and \Planck\footnote{\Planck\
(\url{http://www.esa.int/Planck}) is a project of the European Space Agency
(ESA) with instruments provided by two scientific consortia funded by ESA
member states and led by Principal Investigators from France and Italy,
telescope reflectors provided through a collaboration between ESA and a
scientific consortium led and funded by Denmark, and additional contributions
from NASA (USA).}\ analysis
\citep[][hereafter BKP15]{pb2015}, Galactic foregrounds are known to 
be the dominant component of the $B$-mode polarization signal at high latitudes
and large scales ($\ell < 200$). 
Characterization of these foregrounds is today the main limitation
in the quest for the gravitational wave signature in the $B$-mode cosmic
microwave background (CMB) power spectrum,
i.e., for measuring the tensor-to-scalar ratio, $r$.
This endeavour has been led by the analysis of the Galactic dust foregrounds
carried out using \Planck\ data \citep[][hereafter PIPXXX]{planck2014-XXX}, 
and more recently of the synchrotron foregrounds using \Planck\ and WMAP data
\citep{planck2014-a31,Choi2015,Krachmalnicoff2015}.
It appears that Galactic thermal dust currently represents the major contaminant at
high latitude in the spectral bands mainly adopted to search for the CMB signal by
ground-based and balloon-borne experiments, i.e., between 100 and 220\,GHz \citep{planck2014-XXII,planck2014-a12}.

The efficiency of Galactic dust cleaning for the cosmological $B$-mode analyses
is based on two main factors: the accuracy of the Galactic polarized dust
template; and the way it is extrapolated from the submillimetre to the
millimetre bands.
While recent forecasts of cosmological $B$-modes detection for ground-based,
balloon-borne, and satellite experiments
\citep{Creminelli2015,Errard2015} appear to allow for very good sensitivity
(down to $r\simeq2\times10^{-3}$) with only a few observational bands,
when assuming a simple modelling of the foreground emission, other studies
have shown that even in a global component separation framework, accurate
modelling of the polarized dust spectral energy distribution (SED) is needed in
order to reach the required very low levels of contamination of the cosmological
$B$-modes by Galactic dust residuals
\citep{Armitage-Caplan2012,Remazeilles2015}.
Incorrect modelling of the dust SED
would lead to biased estimates of the $r$ parameter.

Several investigations to quantify the spatial variability of the polarized
dust SED were initiated using the \Planck\ data. A first
estimate of the polarized dust spectral index was discussed
in \citet{planck2014-XXII} for regions at intermediate Galactic latitudes. 
Over 39\,\% of the sky the averaged value has been shown to be slightly larger
than the dust spectral index in intensity, 
with a value of $\beta_{\rm d}^{\rm P}=1.59$, compared with
$\beta_{\rm d}^{\rm T}=1.51$. 
More interestingly, an upper limit of its spatial dispersion was obtained
by computing the standard deviation of the mean polarized dust spectral
index estimated on 352 patches of $10^{\circ}$ diameter, yielding a dispersion
of 0.17.  However, this estimate was shown to be dominated by the expected
\Planck\ noise, and does not allow us to build a reliable model of these spatial
variations.

A second early approach, in PIPXXX, investigated the
correlation ratio between the 217- and 353-GHz \Planck\ bands, which is a
statistical measurement of the dust SED spatial variation, as we will discuss
thoroughly in the following sections. This ratio was computed on large
fractions of the sky and led to an upper limit of 7\,\% decorrelation between
217 and 353\,GHz. 
An initial estimate of the impact of a possible dust polarization decorrelation
between frequencies, due to a spatial variation of its SED, was performed in
the BKP15 analysis. A loss of 10\,\% in the
$150\,{\rm GHz}\times353\,{\rm GHz}$ cross-spectrum in the joint analysis,
due to the decorrelation, was estimated to produce
a positive bias of 0.018 on the determination of $r$. 

In this new study, we present an improved analysis of the correlation ratio
between the 353-GHz and 217-GHz \Planck\ bands as a tracer of the spatial
variations of the polarized dust SED. We derive new constraints on these variations and look at the impact they would have on the determination of the cosmological $B$-mode signal.

This paper is organized as follows.  We first present the data used in this
work in Sect.~\ref{sec:data}.
The analysis of the dust polarization correlation ratio is described in
Sect.~\ref{sec:decorrelation}. The impact of the polarized dust decorrelation on the determination of the
tensor-to-scalar ratio $r$ is illustrated in Sect.~\ref{sec:residuals}.
We discuss the possible origin of the spatial variations of the polarized
dust SED in Sect.~\ref{sec:discussion}, before concluding in
Sect.~\ref{sec:conclusion}.

\section{Data and region selection}
\label{sec:data}

\subsection{Planck data}
\label{sec:planck}

In this paper, we use the publicly available \Planck\
High Frequency Instrument (HFI) data at 217\,GHz and 353\,GHz
\citep{planck2014-a01}. The signal-to-noise ratio of dust polarization in the
\Planck-HFI maps at lower frequencies does not allow us to derive significant
results in the framework adopted for this study and consequently the other
channels are not used. These data consist of a set of maps 
of the Stokes $Q$ and $U$ parameters at each frequency, projected onto the
{\tt HEALPix} pixelization scheme \citep{gorski2005}. The maps and their
properties are described in detail in \citet{planck2014-a09}.

We also use subsets of these data in order to exploit the statistical
independence of the noise between them. As described in \citet{planck2014-a09},
the data were split into either the time or the detector domains. In this paper
we use two particular data splits.

\begin{itemize}

\item The so-called ``detector-set'' maps (hereafter DS, also sometimes called
``DetSets''). \Planck-HFI measures the sky polarization thanks to polarization
sensitive bolometers (PSBs), which are each sensitive to one direction of
polarization. PSBs are assembled in pairs, 
the angle between two PSBs in a pair being 90\deg\ and the angle between
two pairs being 45\deg, allowing for the reconstruction of Stokes $I$, $Q$,
and $U$.  Eight such pairs are available at 217 and 353\,GHz. These were split
into two subsets of four pairs to produce two noise-independent sets of $Q$
and $U$ maps called ``detector-sets,'' for each frequency band.

\item The so-called ``half-mission'' maps (hereafter HM).
\Planck-HFI completed five independent full-sky surveys over 30 months. 
Surveys~1 and 2 constitute the \Planck\ ``nominal mission'' and this was 
repeated a second time during Surveys~3 and 4. A fifth survey was performed
with a different scanning strategy, but was not included in the released
half-mission maps. Thus, \Planck-HFI data 
can be split into two noise-independent sets of $Q$ and $U$ maps,
labelled ``HM1'' and ``HM2,'' for each frequency band.

\end{itemize}

\subsection{Region selection}
\label{sec:masks}

\begin{figure}
\centering
\includegraphics[height=0.25\textwidth]{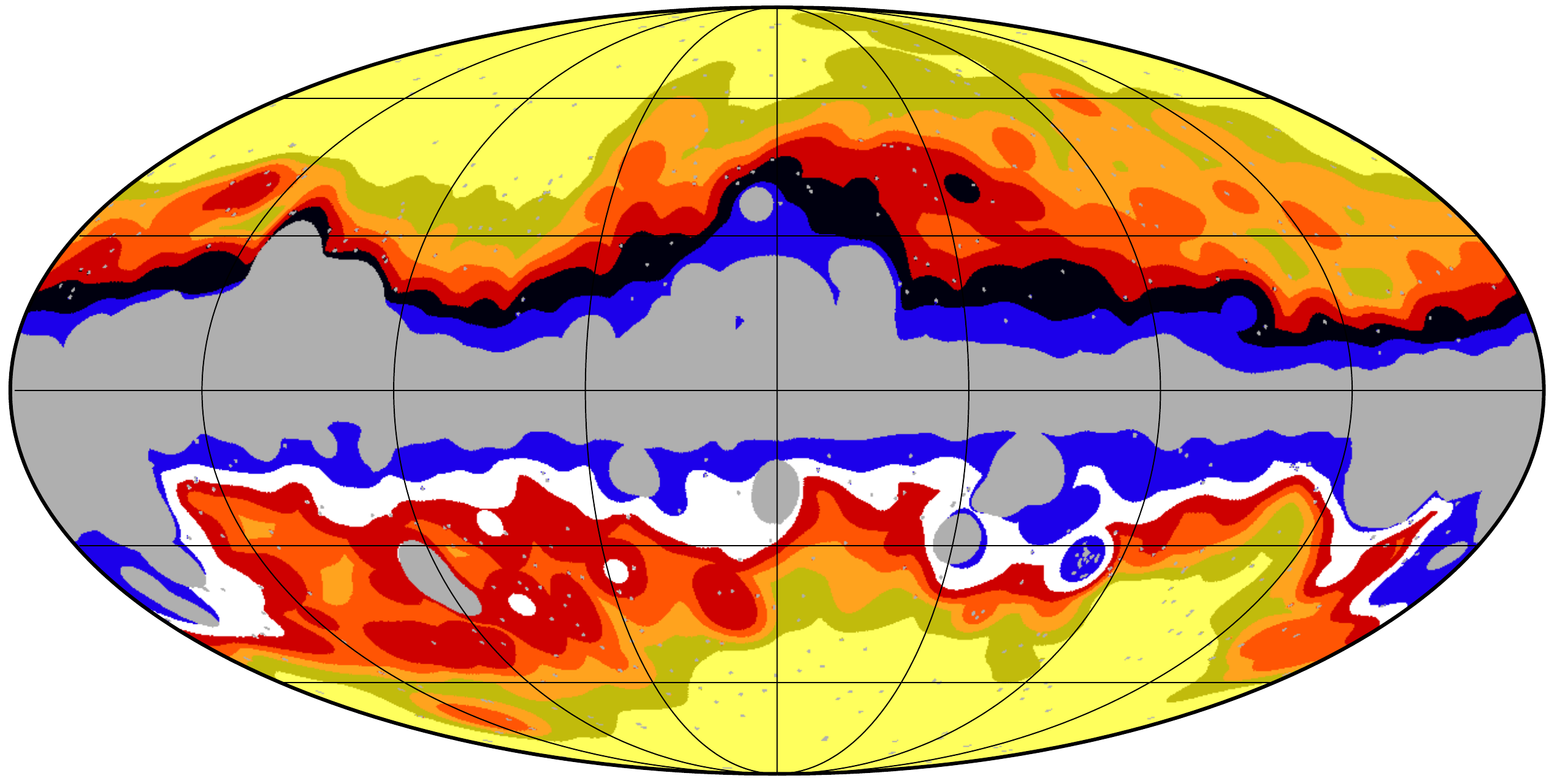}
\caption{
Masks and complementary science regions that
retain fractional coverage of the sky, $f_{\rm sky}$, from 0.8 to 0.2
(see details in Sect.~\ref{sec:masks}).
The grey is the CO mask, whose complement is a selected region with
$f_{\rm sky} = 0.8$.
In increments of $f_{\rm sky}=0.1$, the retained regions can be identified 
by the colours yellow (0.3) to blue (0.8), inclusively. The LR63N and LR63S regions are displayed in black and white, respectively, and the LR63 region is the union of the two. 
Also shown is the (unapodized) point source mask used.
\label{fig:masks}}
\end{figure}

In order to focus on the most diffuse areas of the Galactic dust
emission, we have performed our analysis on various fractions of the
sky using the set of science regions introduced in PIPXXX.  These have
been constructed to reject regions with CO line brightness
$I_{\rm{CO}} \geq 0.4\,\rm{K\,km\,s^{-1}}$.  The complement to this
mask by itself defines a preliminary region that retains a sky
fraction $f_{\rm sky}=0.8$.  In combination with thresholding, based on
the \Planck\ 857\,GHz intensity map, six further preliminary regions
are defined over $f_{\rm sky}=0.2$ to 0.7 in steps of 0.1.  After
point source masking and apodization, this procedure leads to the
large retained (LR) regions LR16, LR24, LR33, LR42, LR53, LR63, and
LR72, where the numbers denote the net effective sky coverages as a
percentage, i.e., $100\, f^{\rm eff}_{\rm sky}.$ We note that the
highest-latitude and smallest region, LR16, is an addition to those
defined in PIPXXX, following the same procedure.  Furthermore, LR63
has been split into its north and south Galactic hemisphere portions,
yielding LR63N and LR63S, covering $f^{\rm eff}_{\rm sky} = 0.33$ and
0.30, respectively. All of these regions are displayed in
Fig.~\ref{fig:masks}.

As previously stated in PIPXXX, the dust polarization angular power
spectra computed on these regions can be considered as approximately
statistically independent because most of the power arises from the
brightest 10\,\% of a given region, the same 10\,\% that
differentiates one (preliminary) region from another.

More specifically, for the $C_\ell^{BB}$ power spectra computed with
the 353-GHz \Planck\ data, the difference (non-overlapping) region
${\rm LR}x_{i} - {\rm LR}x_{i-1}$, $i > 1$ (about 10\,\% of the sky, by
definition), contains more than 75\,\% of the power computed on
LR$x_{i}$, for $x\in\{24,33,42,53,63,72\}$.  LR63N and LR63S are of
course independent of one another, but not of LR63.

We do not use the non-overlapping regions (which would be fully
statistically independent), preferring the LR regions in order to be
able to relate the present work to the previous \Planck\ analyses in general
and to PIPXXX in particular.

\subsection{Simulations}
\label{sec:sims}

The simulated polarization maps presented in this work have been built using a simplified 2-component model
consisting of dust plus CMB, 
both simulated as stochastic realizations with a Gaussian distribution.
CMB maps are defined as realizations based on the Planck Collaboration best-fit $\Lambda$CDM model \citep{planck2014-a15}, 
assuming a tensor-to-scalar ratio $r=0$ and an optical depth $\tau=0.06$ \citep{PlanckReio,PlanckLowell}.
The dust component maps at 353\,GHz are defined as Gaussian realizations using a power-law model of the $EE$ and $BB$ angular power spectra (with a spectral index equal to $-0.42$ and amplitudes matching Table 1 of PIPXXX), 
following the prescriptions of PIPXXX,
and normalized for each region of the sky introduced in Sect.~\ref{sec:masks}. 
Dust maps at other frequencies are scaled with a constant
modified blackbody spectrum with $\beta_{\rm d}^{\rm P}=1.59$ and $T_{\rm d}=19.6$\,K \citep{planck2013-XVII, planck2014-XXII}. 

The instrumental noise component is then introduced in each pixel using the \Planck\ $Q$ and $U$ covariance maps at 217 and 353\,GHz, associated with the DS and HM data setups. 
We checked using a set of 1\,000 simulations that in the range of
multipoles $\ell=50$--700 the instrumental
noise built from covariances was consistent with the  Full Focal Plane
Monte Carlo noise simulations, namely FFP8 \citep{planck2014-a14}
at 217 and 353\,GHz, which are publicly available through the Planck Legacy
Archive.\footnote{\url{http://www.cosmos.esa.int/web/planck/pla}}
For each of the regions described in Sect.~\ref{sec:masks}, we built 1\,000 independent dust, CMB and noise realizations with these properties.

We note that we did not directly use the FFP8 maps in this analysis because of two issues with the polarized dust component
in the \Planck\ Sky Model \citep[PSM, ][]{delabrouille2012} used to build
these simulations.  
First, the polarized dust maps include \Planck\ instrumental noise because they 
are computed from the \Planck\ 353-GHz maps smoothed to 30\arcmin. This contribution significantly enhances the dust signal 
at high Galactic latitudes, where the signal-to-noise ratio of the data is low at this resolution. 
The second issue comes from the fact that the polarized dust component in the
PSM already includes spatial variations of the polarized dust spectral index.
Thus, it cannot be used to perform a null-test for no spatial variation of the SED.

\section{The dust polarization correlation ratio}
\label{sec:decorrelation}

\subsection{Definition}
\label{sec:decorrelationspectra}

We statistically estimate the spatial variation of the dust SED
by constructing the correlation ratio for the dust between 217 and 353\,GHz, as a function of the multipole $\ell$.
This is defined as the ratio of the cross-spectrum between these bands
and the geometric mean of the two auto-spectra in the same bands, i.e.,
\begin{equation}
\label{eq:correlation_ratio}
\mathcal{R}_{\ell}^{XX} \equiv
 \frac{\mathcal{C}^{XX}_{\ell} (353\times217)}
 {\sqrt{\mathcal{C}^{XX}_{\ell}(353\times353) \,
  \mathcal{C}^{XX}_{\ell} (217\times217)}} \, , 
\end{equation}
where $X \in \{E,B\}$. If the maps at 217 and 353\,GHz contain only dust and
the dust SED is constant over the region for which the power spectra are
computed, then $\mathcal{R}_{\ell}^{XX}=1$. 
However, if the dust is not the only component to contribute to the sky
polarization or if the dust SED varies spatially, then
 the ratio is expected to deviate from unity. Nevertheless, the spatial
variations of the SED, as we will see in the next subsection, do not affect the ratio $\mathcal{R}_{\ell}^{XX}$
at the largest scales. 
This is the reason why the study in PIPXXX (where the ratio
$\mathcal{R}_{\ell}^{XX}$ was computed from the fitted 
amplitude of power laws dominated by the largest scales) found no significant
deviation from unity.  In what follows
we conduct this analysis for $\ell > 50$
to avoid any significant contribution from \Planck\ systematics effects
at low $\ell$ \citep{planck2014-a09}.

In order to avoid any bias issues due to the noise auto-correlation, and in
order to minimize the systematic effects, the auto- and cross-spectra
are computed using independent sets of \Planck\ data, namely the half-mission
and detector-set maps \citep{planck2014-a09}.
Hence the auto-spectra are computed for frequency $\nu$ as
\begin{equation}
\mathcal{C}^{XX}_{\ell}(\nu \times \nu)
 \equiv \mathcal{C}^{XX}_{\ell} (D^{1}_{\nu} \times D^{2}_{\nu}) \, ,
\end{equation}
where $D^{1}_{\nu}$ and $D^{2}_{\nu}$ are the two independent sets of data,
e.g., DS1 and DS2, or HM1 and HM2 (see Sect.~\ref{sec:planck}).  Similarly, the cross-spectra between
two frequencies, $\nu_1$ and $\nu_2$, are given by
\begin{equation}
\mathcal{C}^{XX}_{\ell}(\nu_1 \times \nu_2)
 \equiv \frac{1}{4} \sum_{i,j} \mathcal{C}^{XX}_{\ell} (D^{i}_{\nu_1}
 \times D^{j}_{\nu_2}) \, ,
\end{equation}
where the indices $i$ and $j$ take the values 1 and 2.
The results will be labelled ``HM'' or ``DS'' when obtained using the
half-mission or detector-set maps, respectively.

The spectra have been computed using {\tt Xpol} \citep{tristram2005}, which
is a polarization pseudo-$C_{\ell}$ estimator that corrects for incomplete sky
coverage and pixel and beam window functions.

\subsection{\textit{Planck} measurements}
\label{sec:plck_decorrelation_ratio}

\begin{figure}[t]
\begin{tabular}{c}
\includegraphics[width=.5\textwidth]{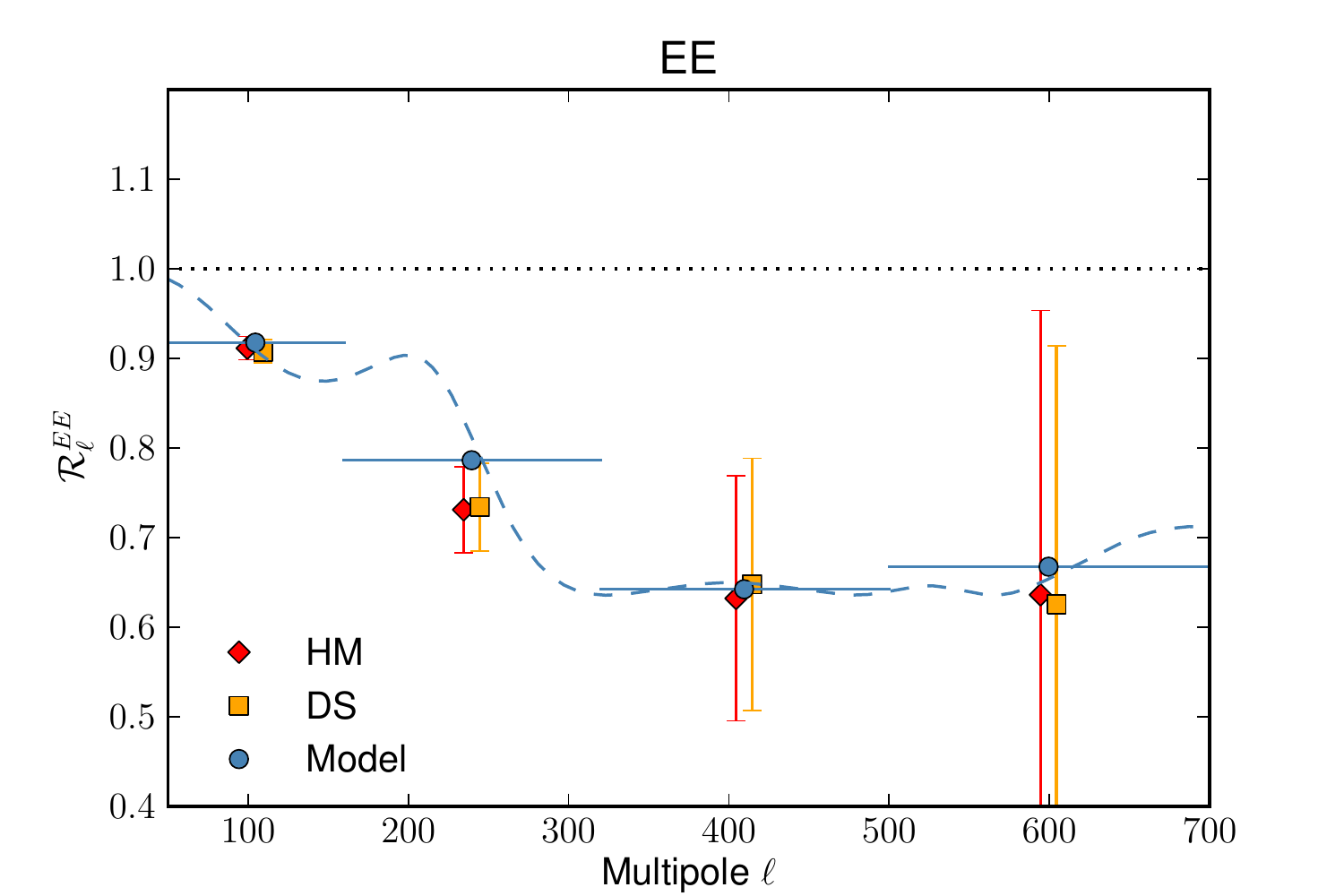} \\
\includegraphics[width=.5\textwidth]{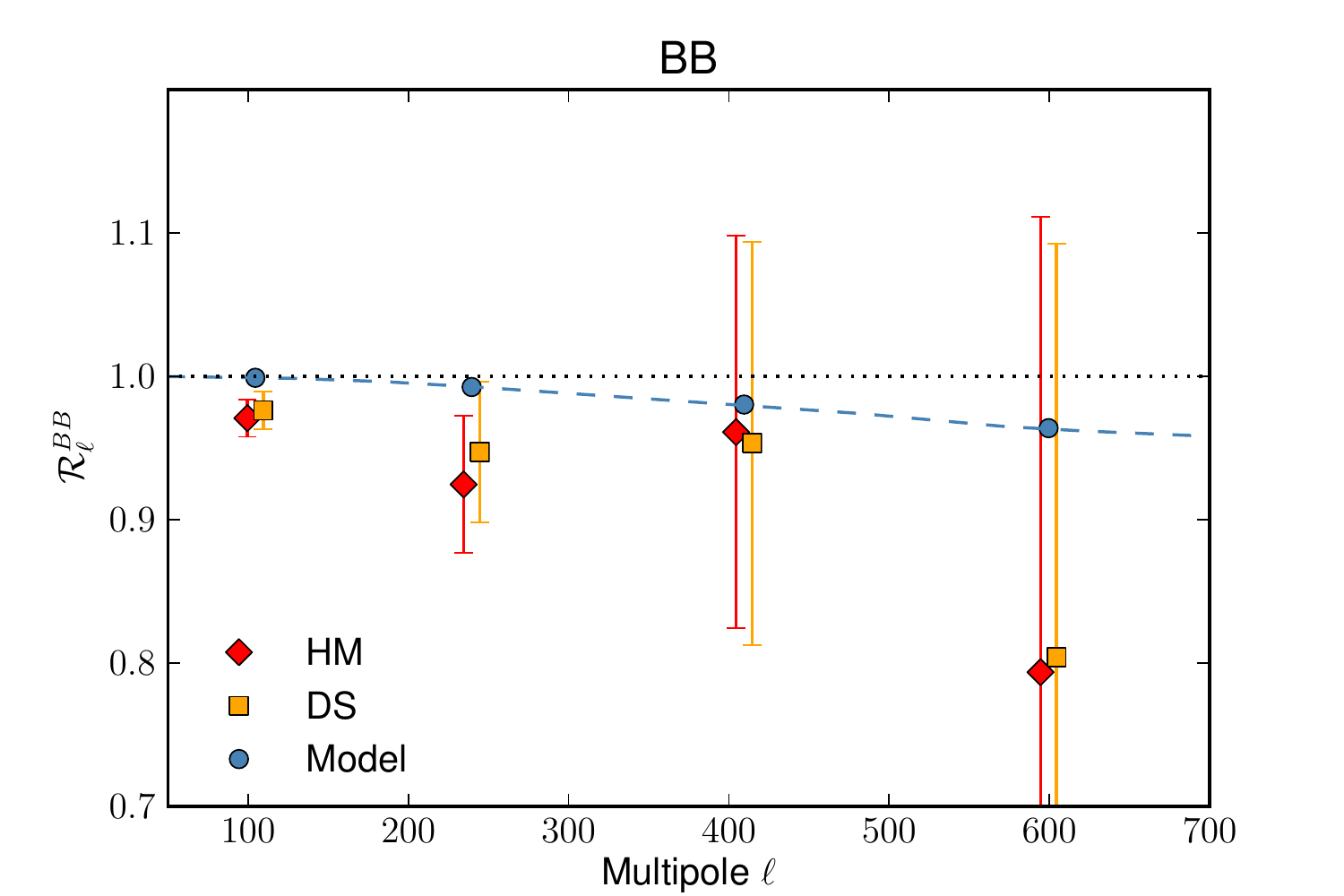} \\
\end{tabular}
\caption{Dust polarization correlation ratios $\mathcal{R}_\ell^{EE}$
(top panel) and $\mathcal{R}_\ell^{BB}$ (bottom panel) between 217 and
353\,GHz, computed on the LR63 region. The correlation ratios determined from
the detector-set data (DS) splits are displayed as yellow squares and the
ratios computed from the half-mission (HM) splits are displayed as red
diamonds. The correlation ratio expectations for the model described in Eq.~(\ref{modelratio})
are displayed as blue dashed lines (blue circles when binned as the data). 
Horizontal blue segments in the top panel represent the range of the $\ell$ bins. 
The uncertainties have been estimated as the median absolute deviation 
over a set of 1\,000 simulations (see Sect.~\ref{sec:sims}) 
of CMB, dust, and Gaussian noise. 
}
\label{fig:powerspectra_EE_BB}
\end{figure}

The $EE$ and $BB$ dust correlation ratios obtained in four multipoles ranges 
($\ell = [50, 160]$, $[160,320]$, $[320,500]$, and $[500,700]$) are shown in
Fig.~\ref{fig:powerspectra_EE_BB} for the \Planck\ HM and DS data versions,
and a fraction of the sky $f_{\rm sky}=0.7$, i.e., the LR63 region.

Since the definition of the correlation ratio, $\mathcal{R}_{\ell}^{XX}$, uses
nonlinear operations (such as ratio and square root),
the associated uncertainties are not trivial to estimate and so are
determined from the median absolute deviation of 1\,000 Monte Carlo 
noise realizations, including \Planck\ instrumental noise, as detailed in Sect.~\ref{sec:sims}.

The correlation ratios between the 217- and 353-GHz bands that might be expected are 
computed using the simulation setup introduced in Sect.~\ref{sec:sims}, and 
defined by a 2-component modelling of dust plus CMB signals, 
assuming no spatial variations of the dust SED and no noise.
If we suppose that the \Planck\ maps at 217 and 353\,GHz are a sum of CMB and
dust components, then (in thermodynamic units and assuming no instrumental
noise) we have
\begin{equation}
M_{353}=M_{\rm dust}+M_{\rm CMB},\nonumber
\end{equation}
\begin{equation}
\label{eq:maps_content}
M_{217}=\alpha M_{\rm dust}+M_{\rm CMB},
\end{equation}
where $M_{\rm dust}$ is the dust map at 353\,GHz, $M_{\rm CMB}$ is the CMB map,
$\alpha$ is a constant scaling coefficient representing the dust SED,
and $M$ can represent the Stokes parameters $Q$ and $U$. Then, combining
Eqs.~(\ref{eq:correlation_ratio}) and (\ref{eq:maps_content})
and assuming that the dust and CMB components are not spatially correlated,
the expected correlation ratio becomes
\begin{equation}
\label{modelratio}
\hat{\mathcal{R}}_{\ell}^{XX} = 
\frac{\alpha \, \mathcal{C}^{XX}_{\ell, \rm{dust}}
 + \mathcal{C}^{XX}_{\ell, \rm{CMB}} }
 {\left[\alpha^2 \left( \mathcal{C}^{XX}_{\ell, \rm{dust}} \right)^2
 + \left( \mathcal{C}^{XX}_{\ell, \rm{CMB}} \right)^2
 + (1+\alpha^2)\,\mathcal{C}^{XX}_{\ell, \rm{dust}}
 \mathcal{C}^{XX}_{\ell, \rm{CMB}}\right]^{1/2} },
\end{equation}
where $X \in \{E,B\}$. It can clearly be seen that even if
$\alpha$ is a constant, the CMB component will make the correlation
ratio $\mathcal{R}\ne1$. The model power spectra corresponding to Eq.~(\ref{modelratio}) are also displayed in Fig.~\ref{fig:powerspectra_EE_BB}.

\begin{table*}[tbp!]
\newdimen\tblskip \tblskip=5pt
\caption{Properties of the nine regions used in this analysis, in terms of effective sky
fraction and column density based on dust opacity (see Sect.~\ref{sec:skyregions}).  The probability to exceed (PTE) values obtained for each multipole bin and sky region
are reported for the HM and DS cases. They are defined as the probability to obtain correlation ratios
smaller than the \Planck\ measurements, based on 1\,000 simulations with dust plus CMB 
signals and Gaussian noise, and expressed as a percentage.}
\label{tab:ptes}
\vskip -6mm
\footnotesize
\setbox\tablebox=\vbox{
 \newdimen\digitwidth
 \setbox0=\hbox{\rm 0}
 \digitwidth=\wd0
 \catcode`*=\active
 \def*{\kern\digitwidth}
  \newdimen\dpwidth
  \setbox0=\hbox{.}
  \dpwidth=\wd0
  \catcode`!=\active
  \def!{\kern\dpwidth}
\halign{\tabskip 0pt#\hfil\tabskip 1em&
\hbox to 3.0cm{#\leaderfil}\tabskip 2em&
\hfil#\hfil\tabskip 0.75em& \hfil#\hfil& \hfil#\hfil&
\hfil#\hfil& \hfil#\hfil& \hfil#\hfil&
\hfil#\hfil& \hfil#\hfil& \hfil#\hfil\tabskip 0em\cr
\noalign{\doubleline}
& \omit & LR16& LR24& LR33& LR42& LR53& LR63N& LR63& LR63S& LR72\cr
\noalign{\vskip 3pt\hrule\vskip 5pt}
\multispan2$f_{\rm sky}^{\rm eff}$ [\%]\hfil& 16& 24& 33& 42& 53& 33& 63& 30& 72\cr
\multispan2$N_{\ion{H}{i}}$ [$10^{20}\,{\rm cm}^{-2}$]\hfil& 1.32& 1.65& 2.12&
 2.69& 3.45& 4.14& 4.41& 4.70& 6.02\cr
\noalign{\vskip 3pt\hrule\vskip 5pt}
& \, $\ell$ range \hfill &\cr
\noalign{\vskip 3pt}
 $\rm{PTE}_{\rm{HM}}$ [\%] &  *50--700 &   *0.0& *0.0&  *0.0&   *0.0&  *0.0&   *0.0& *0.0&  *0.2&   *0.0\cr
 & *50--160 &   *2.5& *2.4&   *2.6&   *1.6&          *0.6&   *6.2&  *1.0&   *5.6&   *0.6\cr
 & 160--320 &   15.0& *5.9&   *1.1&   *6.3&          *7.8&   *2.1&   *7.2&   39.9&   *5.0\cr
& 320--500 &    93.3& 68.5&   59.8&   34.1&          49.4&   36.5&   43.4&   59.2&   40.2\cr
& 500--700 &      \dots& 54.3&   47.0&   30.7&         33.2&   41.4&   27.8&   36.0&   21.4\cr
\noalign{\vskip 3pt\hrule\vskip 5pt}
 $\rm{PTE}_{\rm{DS}}$ [\%] & *50--700 &     *0.0& *0.0&  *1.5&   *0.2&  *0.0&   *0.1& *0.1&  *0.7&   *0.0\cr
 & *50--160 &     *1.7& *1.6&  10.8&   *3.0&  *1.6&   11.2& *3.2&  *6.2&   *4.0\cr
& 160--320 &      29.9& 46.5&   46.4&   42.7&         16.3&   11.5&   14.8&   47.4&   21.7\cr
& 320--500 &   \dots & 84.9&   72.8&   54.4&          42.8&   19.2&   41.9&   73.5&   39.4\cr
& 500--700 &   \dots& \dots&   48.6&   34.6&          36.6&   61.1&   24.5&   18.1&   18.6\cr
\noalign{\vskip 3pt\hrule\vskip 5pt}}}
\endPlancktablewide
\end{table*}

The \Planck\ $EE$ data match the expected $EE$ correlation ratio and are strongly dominated by the
CMB signal. Two approaches have been considered for potentially removing the CMB component from the
correlation ratio in order to see the effect of the dust decorrelation in the $EE$ spectra, namely analysis in either pixel space or multipole space.
The noise on the CMB template, subtracted from the \Planck\
maps, would produce an auto-correlation of the noise when building the 
correlation ratios and would strongly impact our analysis in polarization;
this argues against using the first (pixel-based) option.  Moreover,
 the second option, which consists of correcting the 217- and 353-GHz
\Planck\ cross-spectra by subtracting a model of the CMB power spectrum, 
is affected by the cosmic variance of the CMB, which is dominant
compared to the dust component in the $EE$ correlation ratios.
For all these reasons, in the following analysis, we will focus on the
multipole-based $BB$ modes only, where the CMB component
\citep[coming from the lensing $B$-modes,][]{planck2015-XLI}
is subdominant compared to the observed signal.

As can be seen in the lower panel of Fig.~\ref{fig:powerspectra_EE_BB},
the $BB$ correlation ratios, $\mathcal{R}_{\ell}^{BB}$, 
exhibit a clear deficit compared to the expected model, as discussed in more
detail in the next section.  We have also carried out
similar analyses of the other regions defined in Sect.~\ref{sec:masks}.
The same deficit can be seen
in Fig.~\ref{fig:NH}, where the results averaged over the lowest multipole bin
are shown as a function of the mean column density (see Table~\ref{tab:ptes}).

\subsection{Significance of the \textit{Planck} measurements}
\label{sec:significance}

\begin{figure*}
\includegraphics[width=1.\textwidth]{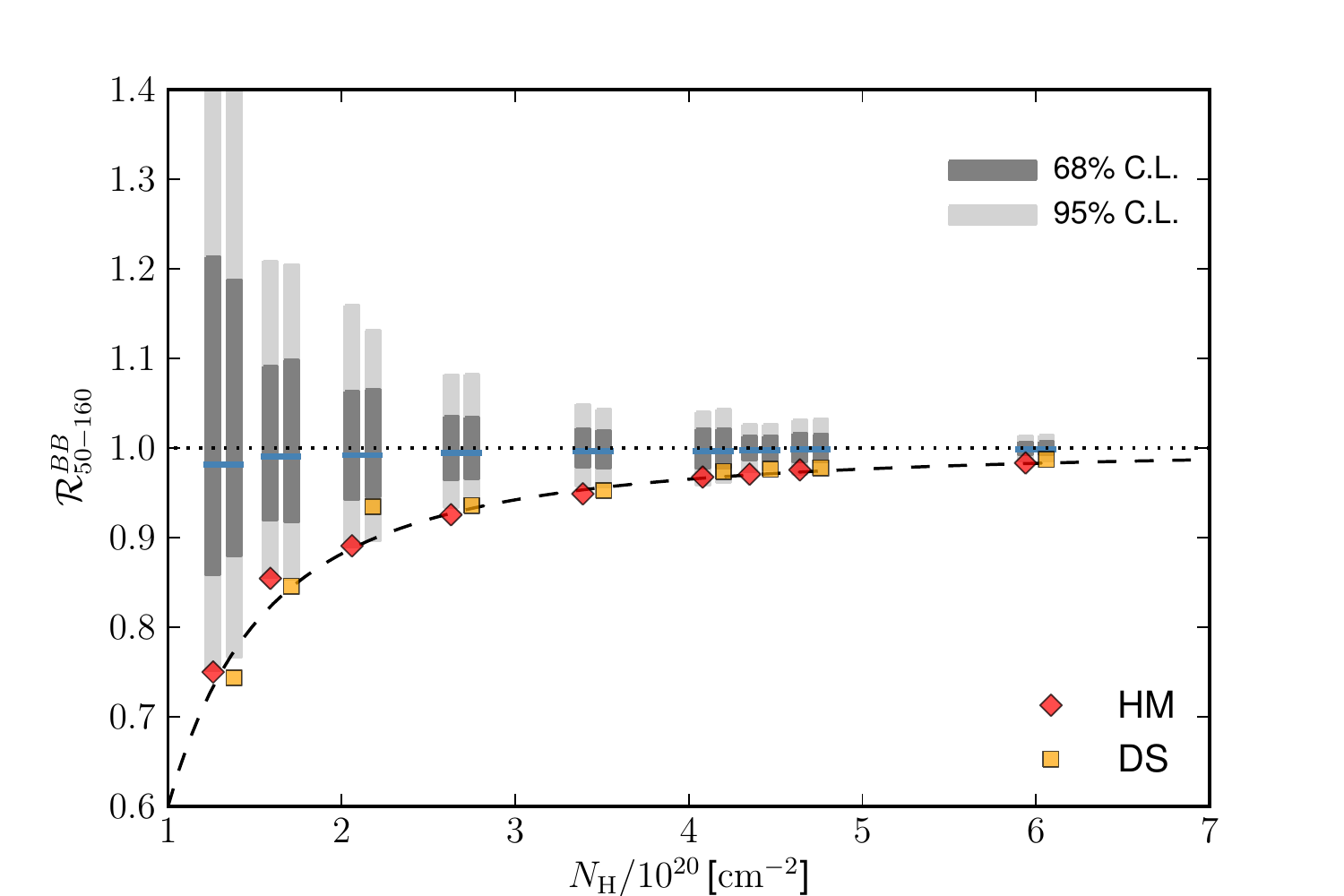} 
\caption{Dependence of the \Planck\ correlation ratios with the mean column density of the region on which they are computed in the first multipole bin ($\ell=[50,160]$). The HM and DS measurements can be compared to the theoretical 
expectations (including dust and CMB, as in Eq.~\ref{modelratio}) shown as blue segments. The grey bars give the 68\,\% and 95\,\% confidence levels computed over 1\,000 realisations of dust, CMB, and Gaussian noise. The column density dependence fit by a power law (see Eq.~\ref{eq:nh_law}) is shown as a dashed line.}
\label{fig:NH}
\end{figure*}

The significance of the \Planck\ data correlation ratios can be quantified using the distribution of $\mathcal{R}_{\ell}^{BB}$ 
computed with simulations including \Planck\ instrumental noise at 217 and 353\,GHz. 
Since the denominator of $\mathcal{R}_{\ell}^{BB}$ can get close to zero, the distribution of the correlation ratios from dust, CMB, and noise can be highly non-Gaussian. For this reason, we cannot give symmetrical error bars and significance levels expressed in terms of $\sigma$. Instead we use the probability to exceed (PTE), which makes no assumption about the shape of the distribution. As we will see, the \Planck\ measurements of the correlation ratios in the DS and HM data sets appear
 systematically smaller than the most probable correlation ratio in our simulations that include instrumental noise.

The impact of the CMB and the noise on the correlation ratio can be seen in Fig.~\ref{fig:NH} 
via the grey vertical histograms. These are based on the correlation ratios obtained on 
a set of 1\,000 Monte Carlo realizations, including
Gaussian \Planck\ noise on top of the simulated CMB and dust components (see Sect.~\ref{sec:sims}).
The distributions of the simulated correlation ratios are available for all multipole ranges and regions in Appendix~\ref{sec:distrib} (in Figs.~\ref{fig:distribs02}--\ref{fig:distribs08}).
While noise barely affects the most probable ratios in the first 
multipole bin ($\ell=50$--160) when compared to the theoretical expectation (blue), 
 it can create an important level of decorrelation in the other bins, particularly in the smallest regions (up to an additional 30\,\% in the fourth multipole bin for LR16). 

In order to quantify the significance of the \Planck\ measurements 
with respect to the simulated decorrelation from CMB and noise, 
we compute the PTE, defined as 
the probability of a simulation having more decorrelation (i.e., smaller correlation ratio) 
than the data. This is computed as
the fraction of the 1\,000 realizations having a correlation ratio
smaller than the \Planck\ measurements, for each multipole bin and HM/DS case. These PTEs are reported in Table~\ref{tab:ptes}.
We also compute a combined PTE over all multipole bins, 
defined as the probability to obtain correlation ratios smaller than \Planck\ measurements simultaneously in all four 
multipole bins ($50 < \ell < 700$).

The combined PTEs (for $50 < \ell < 700$) are $< 1.5\,\%$ for the DS case and $<0.1\%$ for the HM case.
When focusing on individual multipole bins, the detection level is not as strong;
the PTE values range between 0.6\,\% and 11.2\,\% in the first multipole bin ($50<\ell<160$) for both DS and HM cases.
In the second multipole bin  ($160<\ell<320$), the DS correlation ratio PTEs range from 11.5 to 47.4\,\%, while for the HM case they range from 1.1 to 39.9\,\% (with significant PTEs on several regions though).
The third and fourth multipole bins ($320<\ell<700$) show no significant evidence for decorrelation.

The significant excess of decorrelation in the \Planck\ data between 217 and 353\,GHz, especially in the first multipole bin ($50<\ell<160$) is consequently very unlikely to be attributable to CMB or to instrumental noise (or to systematic effects, as discussed in the next section). We therefore conclude that this excess is a statistical measurement of the spatial variation of the polarized dust SED.

\subsection{Impact of systematic effects}
\label{sec:impact_systematics}

Since we have restricted our analysis to multipoles $\ell > 50$, the 217- and
353-GHz cross-spectra are assumed not to be significantly affected by those systematic effects that are most important at low multipoles, such as the 
ADC nonlinearity correction or the dipole and calibration uncertainties
\citep{planck2014-a08, planck2014-a09}.
However, the \Planck\ cross-spectra in the multipole range
$50 < \ell < 700$ could be affected by beam systematics.
Thanks to its definition, the correlation ratio should be approximately
independent of the beam uncertainty, because of
the presence of the same beam functions, $\mathcal{B}_{\ell}^{353}$ and
$\mathcal{B}_{\ell}^{217}$, in the numerator and denominator. 
The only remaining issue could come from the difference
between the beam function of the $353\times217$ cross-spectra,
$\mathcal{B}_{\ell}^{353\times217}$, and the product of the independent beam
functions, $\mathcal{B}_{\ell}^{353} \times \mathcal{B}_{\ell}^{217}$.
We have checked that this ratio exhibits a very low departure from unity,
at the $10^{-5}$ level, which cannot reproduce the
amplitude of the observed \Planck\ correlation ratio.
We also checked that the correction of the bandpass mismatch 
in the 217- and 353-GHz bands does not affect the correlation ratio.
The same analysis has been reproduced using two versions of the
bandpass mismatch corrections \citep[see ][]{planck2014-a08},
yielding results consistent down to 0.1\,\%.

We use the two splits, HM and DS, 
as an indicator of the level of residuals due to systematic effects in our analysis.
This can be assessed by examining Figs.~\ref{fig:distribs02} to \ref{fig:distribs08}. While the HM and DS correlation ratios are very 
consistent in the first and last multipole bins ($\ell=50$--160, and $\ell=500$--700), 
they are in less agreement for the second multipole bin,
($\ell=160$--320). This apparent discrepancy is not explained by the current knowledge of any systematic effects in \Planck, and so indicates the need for some caution.

\subsection{Dependence on column density} 
\label{sec:skyregions}

For CMB polarization studies, it is important to characterize the dependence 
 of the observed decorrelation ratio on column density. 
The \Planck\ correlation ratios in the first multipole bin ($\ell = [50,160]$), obtained 
on the various science regions, are shown in Fig.~\ref{fig:NH}
 as a function of the mean column density, $N_{\ion{H}{i}}$, computed for each region
 as the average over the unmasked pixels of the \Planck\ column density map, assuming a constant opacity
 $\tau / N_{\ion{H}{i}}$ \citep{planck2013-p06b}. 
The HM and DS measurements of the \Planck\ $BB$ correlation ratio 
can be compared to the theoretical expectation (blue segment) and
 the dispersion due to noise (grey histograms) computed over 1\,000 Monte Carlo 
 realizations including Gaussian noise (see Sect.~\ref{sec:sims}).
We recall (Sect.~\ref{sec:masks}) that measurements of the correlation ratio obtained
 in different regions can be considered as statistically independent to a good approximation (except for LR63 with respect to LR63N or LR63S). 
 
In this first multipole bin ($50<\ell<160$), where the primordial $B$-mode signal is expected, 
the $BB$ correlation ratio of the DS and HM cases can be well described 
by a power law of $N_{\ion{H}{i}}$:
\begin{equation}
\label{eq:nh_law}
\mathcal{R}_{50-160}^{BB}  = 1 - K_{50-160}^{BB} \left( \frac{N_{\ion{H}{i}}}{10^{20}} \right)^{\gamma} \, ,
\end{equation}
with $K_{50-160}^{BB}= 0.40 \pm 0.32 $ and $\gamma= - 1.76 \pm 0.69$.
Hence the more diffuse the Galactic foregrounds, the stronger the decorrelation between 217 and 353\,GHz
(this trend is also observed in the last multipole bin with a lower statistical significance, but is less obvious in other bins).
This is an important issue for CMB analyses, which mainly focus on the most diffuse regions of the sky in order to
minimize the contamination by Galactic dust emission.

\section{Impact on the CMB $\boldsymbol{B}$-modes}
\label{sec:residuals}

\begin{figure*}
\includegraphics[width=\textwidth]{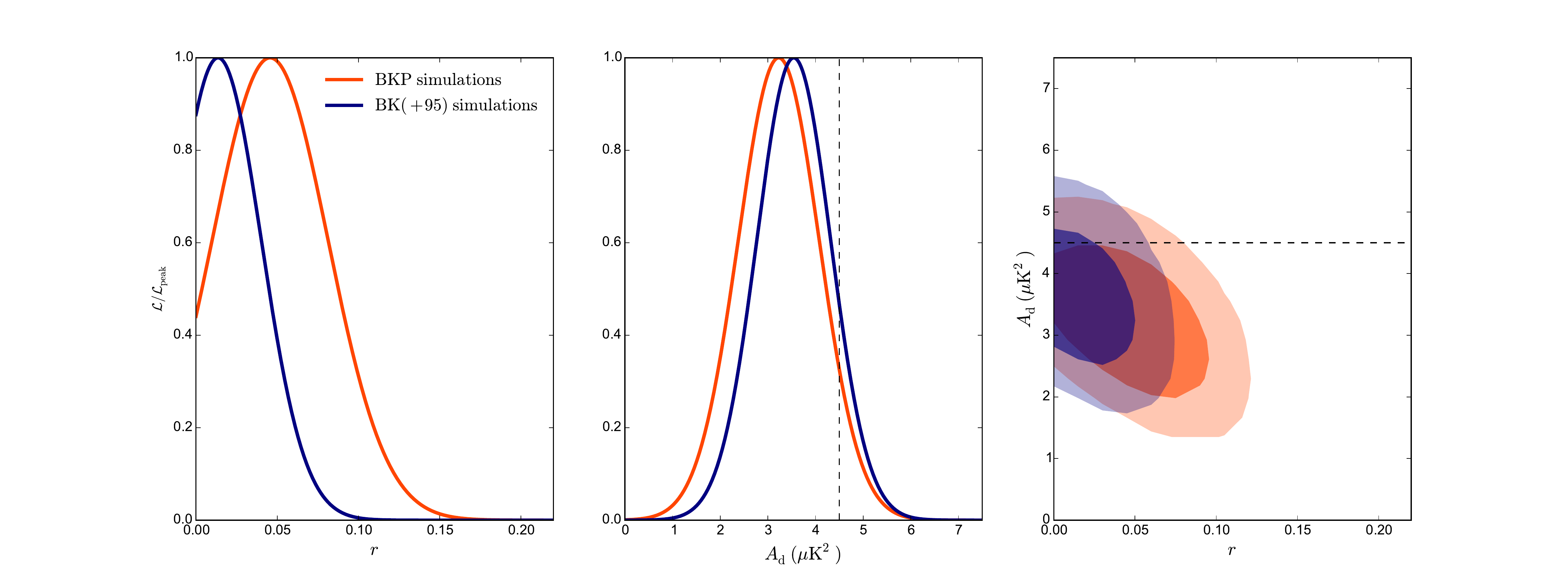} 
\caption{$C_\ell^{BB}$ spectra likelihood posteriors on the $r$ and $A_{\rm d}$ parameters derived from the simulations of the BICEP2-Keck and \Planck\ 353-GHz data, using:
150- and 353-GHz channels (BKP, red); and 95-, 150- and 353-GHz channels (BK(+95), blue).
The 1D posteriors of $r$ (marginalized over $\beta_{\rm d}^{\rm P}$ and
$A_{\rm d}$) and of $A_{\rm d}$ (marginalized over
$\beta_{\rm d}^{\rm P}$ and $r$) are displayed in the left and middle panels
(blue and red lines, respectively). The input value in our simulations for the dust amplitude at 353\,GHz (4.5\,$\mu{\rm K}^2$ at $\ell=80$) is indicated as a dashed line. The 2D posterior
marginalized over $\beta_{\rm d}^{\rm P}$ is presented in the right
panel (68\,\% in darker shading and 95\,\% in lighter shading).}
\label{fig:bkp_likelihood}
\end{figure*}

In Sect.~\ref{sec:decorrelation} we showed that the correlation of the dust
polarization $B$-modes between 217 and 353\,GHz can depart significantly from unity for multipoles $\ell\gtrsim50$. Such a decorrelation, as previously noted by, e.g., \citet{Tassis15} or BKP15, will impact the search for CMB primordial $B$-modes that assume a constant dust polarization SED over the region of sky considered. In order to quantify this effect, we use toy-model simulations of the BICEP2/Keck and \Planck\ data, introducing a decorrelation between the 150- and the 353-GHz channels that matches our results in Sect.~\ref{sec:decorrelation}. This analysis is intended to be illustrative only, and not to be an exact reproduction of the work presented in \citet{pb2015} and \citet{bk2015b}.

We approximate the likelihood analyses presented in BKP15 and \citet{bk2015b} using simple simulations of the CMB and dust $C_\ell^{BB}$ angular power spectra. We directly simulate the $150\times150$, $150\times353$ and $353\times353$ $C_\ell^{BB}$ angular power spectra (where the \Planck\ 353-GHz spectrum comes from two noise-independent detector-set subsamples). In order to include the noise contribution from these experiments and to perform a Monte Carlo analysis over 2\,000 simulations, we add the sample variance and the noise contribution to the spectra as a Gaussian realization of

\begin{equation}
\label{fskyapprow}
\sigma(C_b^{\nu_1\times\nu_2})^2 = \frac{1}{(2\ell+1)f^{\rm eff}_{\rm sky}\Delta\ell_b}\left\{(C_b^{\nu_1\times\nu_2})^2 + C_b^{\nu_1\times\nu_1} C_b^{\nu_2\times\nu_2}\right\}
\end{equation}

\noindent for each power spectra $C_b$ computed in an $\ell$-centred bin of size $\Delta\ell_b$, where $f^{\rm eff}_{\rm sky}$ is the effective sky fraction, $C_b^{\nu_1\times\nu_1}$ and $C_b^{\nu_2\times\nu_2}$ are the signal plus noise auto-power spectra in the frequency bands $\nu_1$ and $\nu_2$, respectively, and $C_b^{\nu_1\times\nu_2}$ is the signal-only cross-power spectrum between these two frequencies (supposing the noise to be uncorrelated between the two bands); notice that while the noise affects the variance of the $C_\ell$, it does not affect their mean value, since we cross-correlate noise-independent maps.

The CMB $C_\ell^{BB}$ spectrum is generated from a \Planck\ 2015 best-fit $\Lambda$CDM model \citep{planck2014-a15} with no tensor modes ($r=0$). The dust for the $353\times353$ power spectrum is constructed as a power law in $\ell$, specifically $\ell^{-0.42}$ following PIPXXX. The amplitude of this spectrum at $\ell=80$ is taken to be $A_{\rm d}=4.5\,\mu{\rm K}^2$. This is an ad hoc value chosen to lie between the predicted PIPXXX value in the BICEP2 region ($A_{\rm d}=13.4\pm0.26\,\mu{\rm K}^2$), and the BKP15 value ($A_{\rm d}=3.3^{+0.9}_{-0.8}\,\mu{\rm K}^2$, marginally compatible with our chosen value, which could be underestimated if some decorrelation exists). A single modified blackbody spectrum is applied to scale the dust 353-GHz $C_\ell^{BB}$ spectrum to the other frequencies, with $\beta^{\rm P}_{\rm d}=1.59$ and $T_{\rm d}=19.6$\,K. 

Finally, we introduce a decorrelation factor $\mathcal{R}_\ell^{BB}$ in our simulated cross-spectra, which we chose to be constant in $\ell$. If we make the assumption that the SED spatial variations come from spatial variations of the dust spectral index around its mean value (see Sect.~\ref{sec:discussionbeta}), a first-order expansion gives a frequency dependence of $(1-\mathcal{R}^{BB})$ that scales as $[\ln(\nu_1/\nu_2)]^2$. We explored many values for the $\mathcal{R}_\ell^{BB}$ ratio and we have chosen to present here the results we obtain for a correlation ratio between 150 and 353\,GHz of $\mathcal{R}_\ell^{BB}(150,353)=0.85$. With the frequency scaling of $[\ln(\nu_1/\nu_2)]^2$, this ratio becomes $\mathcal{R}_\ell^{BB}(217,353)=0.95$, $\mathcal{R}_\ell^{BB}(95,353)=0.65$, and $\mathcal{R}_\ell^{BB}(95,150)=0.96$.

We construct a 3-parameter likelihood function $\mathcal{L}(r,A_{\rm d},\beta^{\rm P}_{\rm d})$, similar to the one used in BKP15, with a Gaussian prior on $\beta^{\rm P}_d=1.59\pm0.11$. For each simulation, the posteriors on $r$ and $A_{\rm d}$ are marginalized over $\beta^{\rm P}_{\rm d}$ and we construct the final posterior as the histogram over 2\,000 simulations of the individual maximum likelihood values for $r$ and $A_{\rm d}$.

Our results when approximating the BKP15 analysis are presented in Fig.~\ref{fig:bkp_likelihood}. The maximum likelihood values are $r=0.046\pm0.036$, or $r<0.12$ at 95\,\% CL, and $A_{\rm d}=3.23\pm0.85\,\mu{\rm K}^2$. The bias on $r$ is higher than the value assessed in section~V.A of BKP15 (we tested that our simulations give the same 0.018 bias on $r$ when using $\mathcal{R}_\ell^{BB}(150,353)=0.90$), given that we introduce more decorrelation and that our dust amplitude is higher. The value we find for $A_{\rm d}$ is similar to that of the BKP15 analysis.

Finally, we repeat the same analysis on simulations corresponding to \citet{bk2015b}, where we add the 95-GHz data and increase the sensitivity in the 150-GHz channel with respect to BKP15. Unlike in \citet{bk2015b}, we do not parametrize a synchrotron component. The posteriors from these simulations, labelled ``BK(+95),'' are also displayed in Fig.~\ref{fig:bkp_likelihood}. The maximum likelihood values are $r=0.014\pm0.027$ (or $r<0.07$ at 95\,\% CL) and $A_{\rm d}=3.54\pm0.77\,\mu{\rm K}^2$. Even without a synchrotron component in the model of the data, the positions of the peak in the posteriors on $r$ and $A_{\rm d}$ with respect to BKP15 are shifted in the same direction as found in \citet{bk2015b}.

The region of the sky observed by the BICEP and Keck instruments presented in BKP15 has a mean column density of $N_{\ion{H}{i}}=1.6\times10^{20}\,{\rm cm}^{-2}$, very similar to the one for our region LR24. In the multipole range $50<\ell<160$, using the empirical relation derived in Sect.~\ref{sec:skyregions} (Eq.~\ref{eq:nh_law}), we expect to have a decorrelation $\mathcal{R}_{50-160}^{BB}(217,353)=0.85$ between 217 and 353\,GHz. Introducing the latter decorrelation in our simple simulations shifts the $r$ posterior towards higher values, making a decorrelation as high as the one we measure on the LR24 mask very unlikely, given the BKP15 data. This shows the limitation of the empirical relation we derived in Sect.~\ref{sec:skyregions} when dealing with small regions, since the properties of the decorrelation might be very variable over the sky. A specific analysis of these data could quantitatively confirm the amount of decorrelation that is already allowed or excluded by the data. 

The results presented in this section stress that a decorrelation between the dust polarization at any two frequencies will result in a positive bias in the $r$ posterior, in the absence of an appropriate modelling in the likelihood parametrization or in any component separation. The current BICEP2/Keck and \Planck\ limits on $r$ still leave room for a decorrelation of the dust polarization among frequencies that could be enough to lead to spurious detections for future Stage-III or Stage-IV CMB experiments \cite[see, e.g.,][]{Wu2014, Errard2016}. 

\section{Discussion}
\label{sec:discussion}

We now quantify how the observed decorrelation of the $BB$ power spectrum 
between the 217- and 353-GHz bands can be explained by spatial variations of the polarized dust SED using two toy-models
presented in Sects.~\ref{sec:discussionbeta} and \ref{sec:discussionangles}.
This simplified description in characterizing spatial variations of the dust SED is a first step, which ignores correlations between dust properties and the structure of the
magnetized interstellar medium.  Correlations between matter and the Galactic magnetic field have been shown to be essential to account for statistical properties of dust 
polarization at high Galactic latitudes \citep{Clark15,planck2015-XXXVIII}.
In Sect.~\ref{sec:origin}, we use the framework introduced by \citet{planck2016-XLIV}
to discuss why such correlations are also likely to be an essential
element of any physical account of the variations of the dust SED in polarization.

\subsection{Spectral index variations}
\label{sec:discussionbeta}

In a first approach, we assume that the variations of the polarized dust SED can
be fully explained by spatial variations of the polarized dust spectral index applied
simultaneously to the Stokes $Q$ and $U$ components.

We make a simplifying approximation by assuming that the polarized dust spectral index
follows a Gaussian distribution centred on the mean value $\overline{\beta_{\rm{d}}^{\rm{P}}} = 1.59$ \citep{planck2014-XXII}, 
with a single dispersion, $\Delta \delta_{\rm{d}}^{\rm{P}}$, at all scales over the
whole sky.  Specifically the dust polarization spectral index is given by
\begin{equation}
\beta_{\rm{d}}^{\rm{P}} \left( \hatn\right) = \mathcal{N}
 \left( \overline{\beta_{\rm{d}}^{\rm{P}}}, \Delta \delta_{\rm{d}}^{\rm{P}}
  \right) \left( \hatn\right),
\end{equation}
where $\mathcal{N}(x_0, \sigma)$ is a Gaussian distribution centred on $x_0$
with a standard deviation $\sigma$, and $\Delta \delta_{\rm{d}}^{\rm{P}}$
is the dispersion of the spectral index map, defined 
as the standard deviation after smoothing at a resolution of $1^{\circ}$.
The dust temperature is kept constant over the sky
 and equal to $T_{\rm{d}}=19.6\,\rm{K}$ \citep{planck2014-XXII}.

A main caveat of this approach comes from the power
 introduced by the spatial variations of the spectral index, 
 which alters the 
 power spectrum of dust polarization.
 This effect remains small between 217 and 353\,GHz for low values of $\Delta \delta_{\rm{d}}^{\rm{P}}$ at $\ell < 150$, 
 but can lead to dust power spectra that are inconsistent with \Planck\ observations when extrapolated to further bands.
This effect is inherent to this modelling approach, which must only be considered as illustrative. However, here the simulated maps at 217 and 353\,GHz are built from maps at an intermediate frequency ($\sqrt{217\times353}\simeq 277$\,GHz) to minimize the addition of power.

The $BB$ correlation ratio model between 217 and 353\,GHz is constructed as follows.
We start with a set of $Q$ and $U$ dust template maps at 277\,GHz, appropriately normalized to \Planck\ data as detailed in Sect.~\ref{sec:sims}. 
The polarization dust maps at 217 and 353\,GHz are extrapolated from 277\,GHz 
using a Gaussian realization of the polarized dust spectral index, 
given a level of the dispersion $\Delta \delta_{\rm{d}}^{\rm{P}}$.
These simulated maps do not include noise at this point.
The correlation ratio model is finally obtained by averaging 100 realizations
of the correlation ratio computed on a pair of simulated maps.

\begin{figure}
\center
\includegraphics[width=0.5 \textwidth]{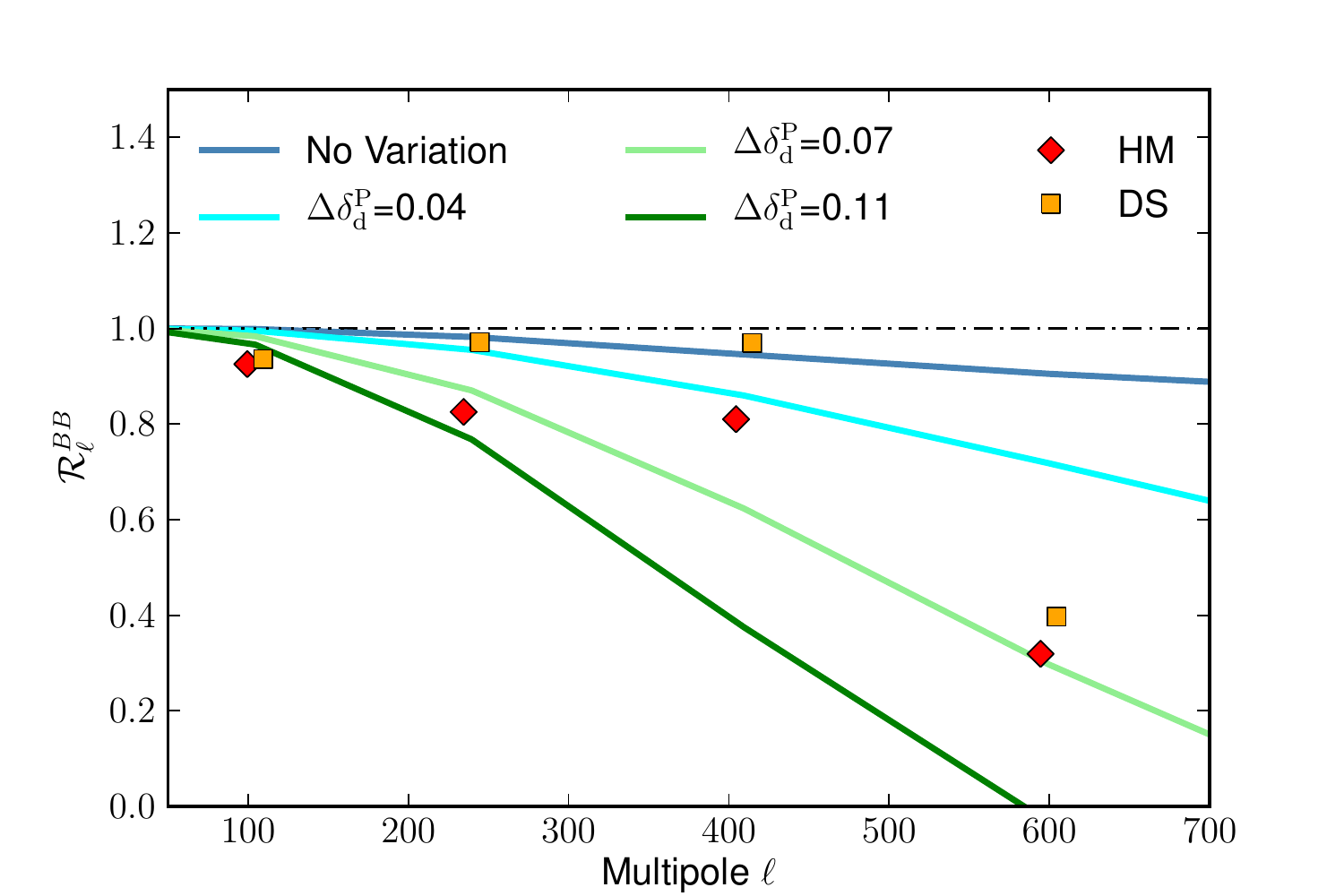} 
\caption{Illustration of the correlation ratio modelled with Gaussian spatial variations of the polarized dust spectral index,
 in LR42 as in Fig.~\ref{fig:powerspectra_EE_BB}. The HM and DS \Planck\ measurements 
are shown as diamonds and squares, respectively.
The model is plotted for four values of $\Delta \delta_{\rm{d}}^{\rm{P}}$, which is
 defined as the standard deviation of the Gaussian realization of the spectral index after smoothing at $1\deg$. The first model, with $\Delta \delta_{\rm{d}}^{\rm{P}}=0$, is that described in Sect.~\ref{sec:plck_decorrelation_ratio}.}
\label{fig:models_GC}
\end{figure}

This model is illustrated in Fig.~\ref{fig:models_GC} for three non-zero values of 
$\Delta \delta_{\rm{d}}^{\rm{P}}$ and compared to the \Planck\
HM and DS measurements in the LR42 region.
In order to match the \Planck\ data (including uncertainties), an
indicative value of $\Delta \delta_{\rm{d}}^{\rm{P}}$ 
around 0.07 is suggested by this simple analysis.

We compare our estimate of the spectral index variations for dust polarization to those measured for the total dust intensity. 
We use the {\tt Commander} \citep{eriksen2006, eriksen2008} dust component maps
derived from a modified blackbody fit to \Planck\ data \citep{planck2014-a12}, providing two separate maps of
dust temperature and dust spectral index. From the ratio $I_{353}/I_{217}$ computed from these two maps, we derive an equivalent all-sky map
of the intensity dust spectral index, $\tilde{\beta}_{\rm d}^{\rm I}$, assuming a constant dust temperature of $T_{\rm d}=19.6\,{\rm K}$.
We use the two half-mission maps of the dust spectral index, computed from 
$1\deg$ resolution maps, instead of the full-survey map, in order to reduce the impact
of noise and systematic effects when computing the covariance,
and we derive an estimate of the standard deviation of the dust spectral index in intensity,
$\Delta \delta_{\rm{d}}^{\rm{I}} \approx 0.045$ in LR42, about half the value 
measured for polarization in the same region. We note that this value is not corrected for the contribution of the
data noise and the anisotropies of the cosmic infrared background \citep{planck2014-XXII}. Both of these contribution are much smaller than the empirical value of 0.17 found in \citet{planck2014-XXII}, which is dominated by noise.

\subsection{Polarization angle variations}
\label{sec:discussionangles}

In a second approach, we assume that the decorrelation of the $BB$ power
spectrum between 217 and 353\,GHz can be explained by spatial variations of
the polarization angle, keeping the dust temperature and polarized spectral
index constant over the whole sky. Unlike the first model, this modelling approach conserves the
total power in the power spectrum.
It is motivated by the nature of the polarized signal, which can be
considered as the sum of spin-2 quantities over multiple components with varying spectral dependencies along the line of sight. Physical interpretation of polarization angle variations are further discussed in Sect.~\ref{sec:origin}.

The spatial variations of the polarization angle are assumed to follow a
circular normal distribution (or von Mises distribution) around 0, given by
\begin{equation}
f(\theta \, | \, \kappa) = \frac{e^{\kappa \cos (\theta)}}{2\pi \mathcal{I}_0(\kappa)} \, ,
\end{equation}
where $\mathcal{I}_0(x)$ is the modified Bessel function of order 0, and $\kappa$ is 
analogous to $1/\sigma^2$ for the normal distribution. 
 While the circular normal distribution allows us to define random angles in the range $[-\pi, \pi]$, 
we re-scale the realizations of angular variations
to match the definition for the range of polarization angles lying between $-\pi/2$ and $\pi/2$.
Again we start from a set of $Q$ and $U$ dust maps at 277\,GHz, 
which are then extrapolated to 217 and 353\,GHz following a 
modified blackbody spectrum using $T_{\rm d}=19.6$\,K and $\beta_{\rm{d}}^{\rm{P}}=1.59$. 
The polarization pseudo-vector obtained from $Q$ and $U$ is rotated independently at each frequency 
using two different realizations of the polarization angle variations.
The correlation ratio model is then obtained by averaging 100 realizations of the 
correlation ratio computed from a pair of simulated maps.

\begin{figure}
\center
\includegraphics[width=0.5 \textwidth]{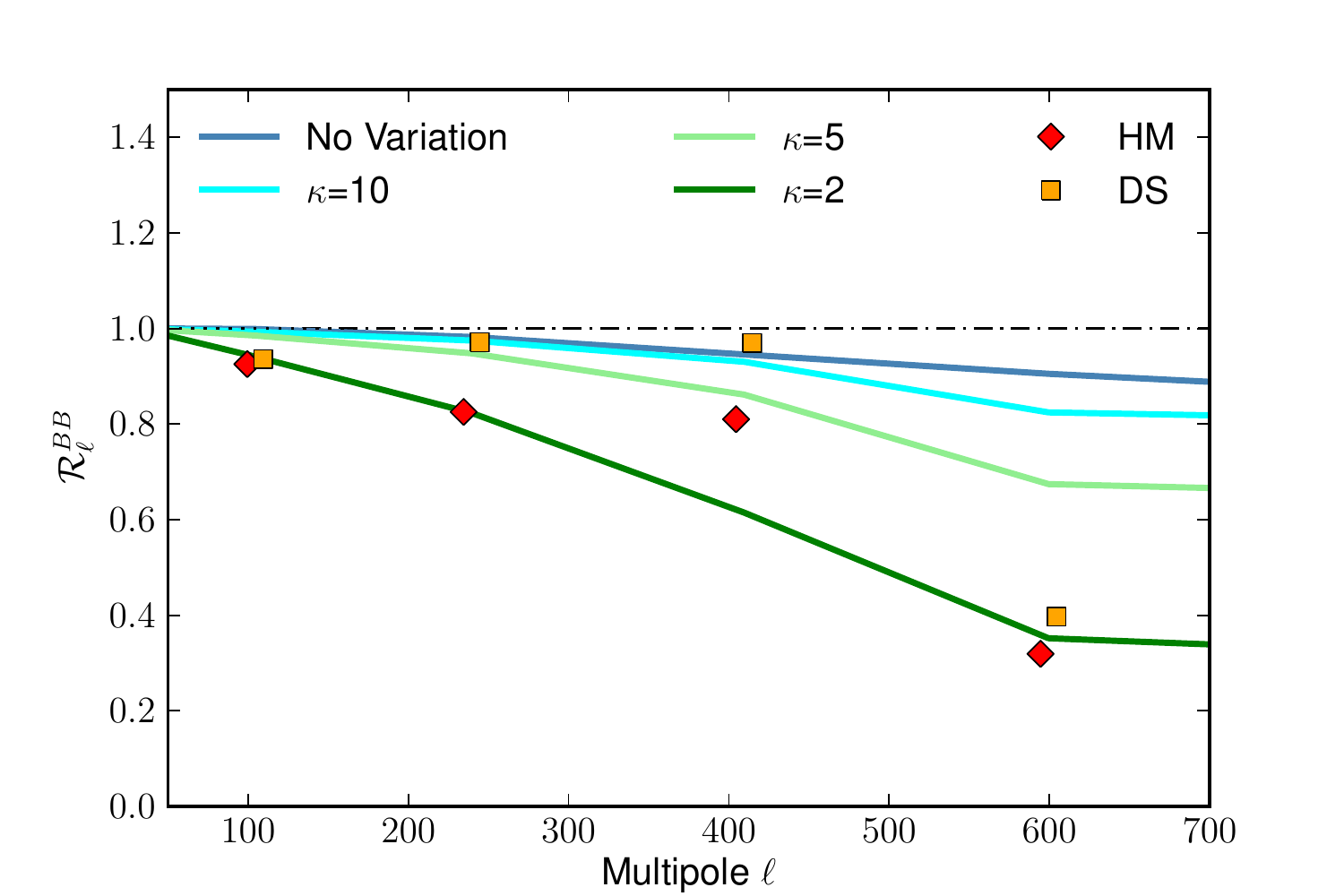} 
\caption{Like Fig.~\ref{fig:models_GC}, but illustrating the correlation ratio modelled with Gaussian spatial variations of the polarization angle.
The model is plotted for four values of $\kappa$, which sets the level of the circular normal distribution for polarization angle, analogous to $1/\sigma^2$ for the normal distribution.}
\label{fig:models_GA}
\end{figure}

This model is illustrated Fig.~\ref{fig:models_GA} for three finite values of 
$\kappa$ in the LR42 region. 
The case $\kappa=2$, which matches the \Planck\ data quite well, 
represents a $1\,\sigma$ dispersion of the polarization angle of
about $2^{\circ}$ between the 217- and 353-GHz polarization maps,
after smoothing at $1^{\circ}$ resolution.

\subsection{Origin on the spatial variations of the polarized SED}
\label{sec:origin}

\begin{figure}
\center
\includegraphics[width=0.45\textwidth]{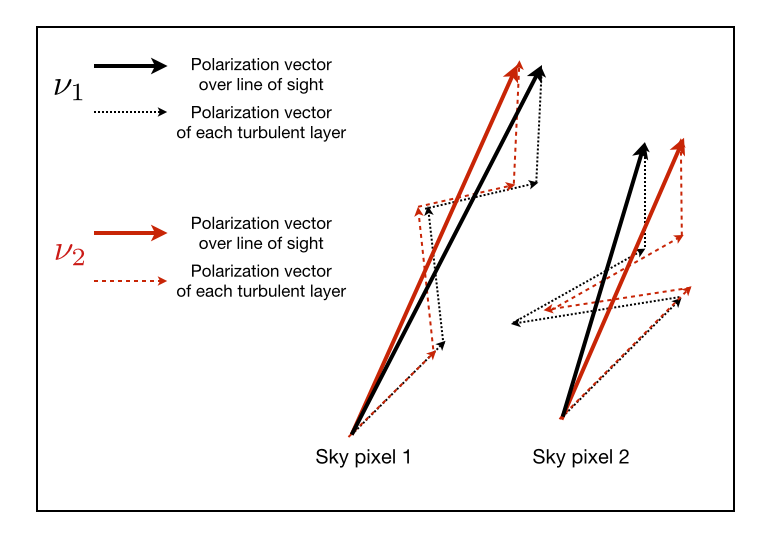} 
\caption{Illustration of the decomposition of the line of sight polarization
pseudo-vector into a random walk process through four turbulent layers. This is
shown for two neighbouring pixels at two frequencies, $\nu_1$ and $\nu_2$.
The integrated polarization pseudo-vectors are affected by the polarization
angle fluctuations, leading to a decorrelation between the two frequencies.}
\label{fig:randomwalk}
\end{figure}

The spatial variations of the Galactic dust SED are found to be larger for
dust polarization than for dust intensity. 
This difference could result from the lack of correlation 
between spectral variation and emission features in our modelling in Sect.~\ref{sec:discussionbeta}.
It could also reflect the different nature of the two observables. While
variations in the dust SED tend to average out in intensity (because it is a
scalar quantity), they may not average out as much in polarization (because it
is a pseudo-vector). In other words, dust polarization depends on the magnetic
field structure, while dust intensity does not. In this section, we
describe further details of this interpretation of the decorrelation.

The analysis of \Planck\ data offers several lines of evidence for the imprint
of interstellar magneto-hydrodynamical (MHD) turbulence on the dust
polarization sky.  Firstly, \citet{planck2014-XIX} reported an anti-correlation
between the polarization fraction and the local dispersion of the polarization
angle.  Secondly, large filamentary depolarization patterns observed in the
\Planck\ 353-GHz maps are associated, in most cases, with large, local,
fluctuations of the polarization angles related to the magnetic field
structure \citep{planck2014-XIX}.
Thirdly, the polarization fraction of the \Planck\ 353-GHz map shows a large
scatter at high and intermediate latitudes, which can be interpreted
as line of sight depolarization associated with interstellar MHD turbulence
\citep{planck2014-XX,planck2016-XLIV}.
Lastly, several studies have reported a clear
trend where the magnetic field is locally aligned 
with the filamentary structure of interstellar matter \citep{Martin2015,Clark15,planck2014-XXXII,planck2015-XXXVIII,Kalberla16}.

As presented in \citet{planck2016-XLIV}, the \Planck\ polarized data
at high Galactic latitudes can be 
modelled using a small number of effective polarization 
layers in each of which the Galactic magnetic field has a turbulent component of the magnetic field 
with a distinct orientation. Within this model, the integration of dust
polarization along the line of sight can be viewed as the result of an oriented random walk
in the ($Q$, $U$) plane, with a small number of steps tending towards the mean direction of the Galactic magnetic field.
The magnetic field orientation of each turbulent layer sets the direction of
the step, while the dust polarization properties, including the efficiency of
grain alignment, sets their length. Changes in dust properties generate
differences in the relative lengths of the steps across frequency, and thereby
differences in the polarization fraction and angle, as illustrated graphically
in Fig.~\ref{fig:randomwalk}.
In this framework, variations of the dust SED along the line of sight impact
both the polarization fraction and the polarization angle, because
the structure of the magnetic field and of diffuse interstellar matter are
correlated \citep{planck2014-XXXII,planck2015-XXXVIII}. 
In the random walk, the variations in length and angle of the polarization
pseudo-vector have most impact for a small number of steps.
\citet{planck2016-XLIV} related the number of layers in their model 
to the density structure of the diffuse interstellar medium 
and to the correlation length of the turbulent component of the magnetic field along the line of sight.

A quantitative modelling of this perspective on dust polarization is required
to assess its ability to reproduce the results of our analysis of the \Planck\
data and its impact on the dust/CMB component-separation task.
To our knowledge, none of the studies carried out so far to quantify limits
set by polarized dust foregrounds on future CMB experiments in order to
search for primordial $B$-modes \citep[except for the work of][]{Tassis15}
have considered the frequency decorrelation that arises from the interplay
between dust properties and interstellar MHD turbulence.

\section{Conclusions}
\label{sec:conclusion}

We have used the 217- and 353-GHz \Planck\ 2015 data to investigate
the spatial variation of the polarized dust SED at high Galactic latitudes.
We computed $\mathcal{R}_{\ell}^{BB}$---the $BB$ cross-spectrum between the
\Planck\ polarization maps in these bands divided by the geometric mean of
the two auto-spectra in the same bands---over large regions of the sky at
high latitudes, for multipoles $50< \ell < 700$.
$\mathcal{R}_{\ell}^{BB}$ was computed with distinct sets of data to control systematics.

The ratio  $\mathcal{R}_{\ell}^{BB}$ has been shown to be significantly lower than what is expected purely from the presence of CMB and noise in the data, with a confidence,
estimated using data simulations, larger than 99\,\%. We interpret this result as evidence for significant spatial variations of the dust polarization SED. 

In the multipole bin $50 < \ell < 160$ that encompasses the recombination bump of the primordial $B$-mode signal, 
$\mathcal{R}_{\ell}^{BB}$ values are consistent for the distinct \Planck\ data splits we used. 
The measured values exhibit a systematic trend with
 column density, where $\mathcal{R}_{\ell}^{BB} $ decreases for decreasing mean column densities $N_{\ion{H}{i}}$
 as $ 1 - K_{50-160}^{BB}(N_{\ion{H}{i}}/10^{20}\,{\rm cm}^{-2})^{\gamma}$,
 with $\gamma=-1.76\pm0.69$ and $K_{50-160}^{BB}= 0.40 \pm 0.32$. This suggests that, statistically speaking, the cleaner a
sky area is from Galactic foregrounds, the more challenging it may be to extrapolate dust polarization from submm
to CMB bands. 

The spatial variations of the dust SED are shown to be stronger in polarization
than in intensity.  This difference may reflect the interplay, encoded in the
data, between the polarization properties of dust grains, including grain
alignment, the Galactic magnetic field, and the density structure of
interstellar matter.

We have proposed two toy-models to quantify the observed decorrelation, based
on spatial Gaussian variations of the polarized dust spectral index, and
spatial Gaussian variations of the polarization angle.  Both models reproduce
the general trend observed in \Planck\ data, with reasonable accuracy given
the noise level of the \Planck\ measurements.  They represent a first step
in characterizing variations of the dust SED, as a prelude to further work on
a physically motivated model, which will take into account the expected
correlation between the spatial variations of the SED and structures in the
dust polarization maps.

Spatial variations of the dust SED can lead to biased estimates of the
tensor-to-scalar ratio, $r$, because of inaccurate extrapolation of dust
polarization when cleaning or modelling the $B$-mode signal at microwave
frequencies.  As an illustration, we have shown that in a region such as that
observed by the BICEP2-Keck Array, a decorrelation of 15\,\% between the 353-
and 150-GHz bands could result in a likelihood posterior on
the tensor-to-scalar ratio of $r=0.046\pm0.036$, comparable to the
joint BICEP2 and Keck Array/Planck Collaboration result.

It appears essential now to place tighter constrains on the spectral dependence
of polarized dust emission in the submm, in order to properly propagate the
information on the Galactic foregrounds into the CMB bands.  More
specifically, the spatial variations of the polarized dust SED need
to be mapped and modelled with improved accuracy in order to be able to
confidently reach a level of Galactic dust residual lower than $r\sim10^{-2}$.

\begin{acknowledgements}
The Planck Collaboration acknowledges the support of: ESA; CNES, and
CNRS/INSU-IN2P3-INP (France); ASI, CNR, and INAF (Italy); NASA and DoE
(USA); STFC and UKSA (UK); CSIC, MINECO, JA, and RES (Spain); Tekes, AoF,
and CSC (Finland); DLR and MPG (Germany); CSA (Canada); DTU Space
(Denmark); SER/SSO (Switzerland); RCN (Norway); SFI (Ireland);
FCT/MCTES (Portugal); ERC and PRACE (EU). A description of the Planck
Collaboration and a list of its members, indicating which technical
or scientific activities they have been involved in, can be found at
\href{http://www.cosmos.esa.int/web/planck/planck-collaboration}{\texttt{http://www.cosmos.esa.int/web/planck/planck-collaboration}}.
The research leading to these results has received funding from the European
Research Council under the European Union's Seventh Framework Programme
(FP7/2007-2013) / ERC grant agreement No.~267934.
\end{acknowledgements}

\bibliographystyle{aat}

\bibliography{PIP_L_Montier_Aumont.bbl}


\appendix

\section{Distribution of the $\mathcal{R}_\ell^{BB}$ from \Planck\ simulations.}
\label{sec:distrib}

We present in this appendix the distribution of the ratio $\mathcal{R}_\ell^{BB}$ for our simulations of the \Planck\ 217- and 353-GHz maps, including Gaussian CMB, dust, and \Planck\ noise, which are described in Sect.~\ref{sec:significance}. Since the quantity $\mathcal{R}_\ell^{BB}$ is a ratio, we expect distributions that are strongly non-Gaussian, even for Gaussian simulations, as soon as the denominator gets close to zero. 

\begin{figure}
\includegraphics[width=0.5\textwidth]{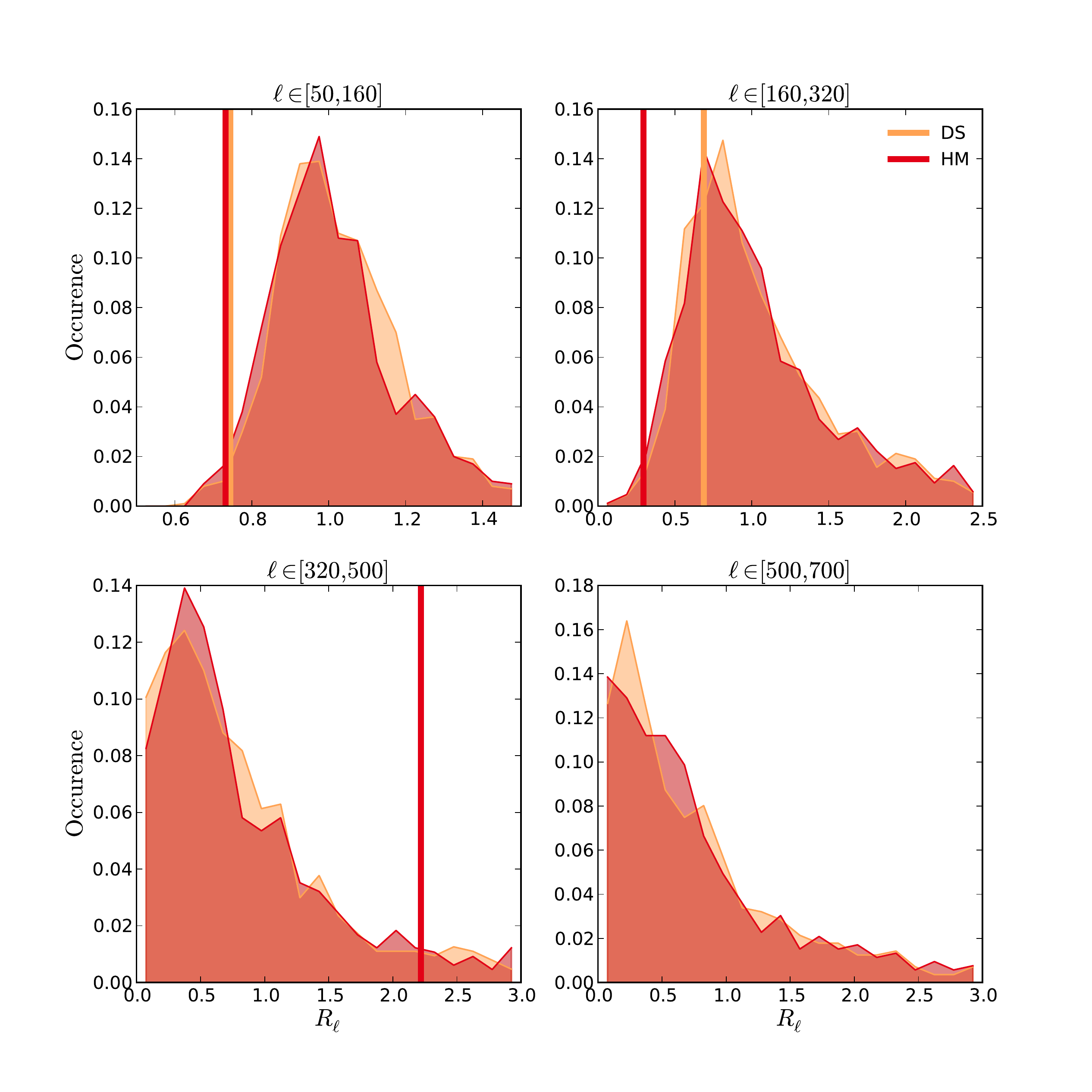} 
\caption{Distribution of the correlation ratios $\mathcal{R}_\ell^{BB}$ on the LR16 region from \Planck\ simulations for detector-set (DS, red histograms) and half-mission (HM, orange histograms) data splits, for the four multipole bins we use in the analysis: top left, $\ell\in[50,160]$; top right, $\ell\in[160,320]$; bottom left, $\ell\in[320,500]$; and bottom right, $\ell\in[500,700]$. The values derived from the data are displayed as vertical lines of the corresponding colour. The PTEs corresponding to these data values with respect to the simulations are reported in Table~\ref{tab:ptes}.}
\label{fig:distribs02}
\end{figure}

\begin{figure}
\includegraphics[width=0.5\textwidth]{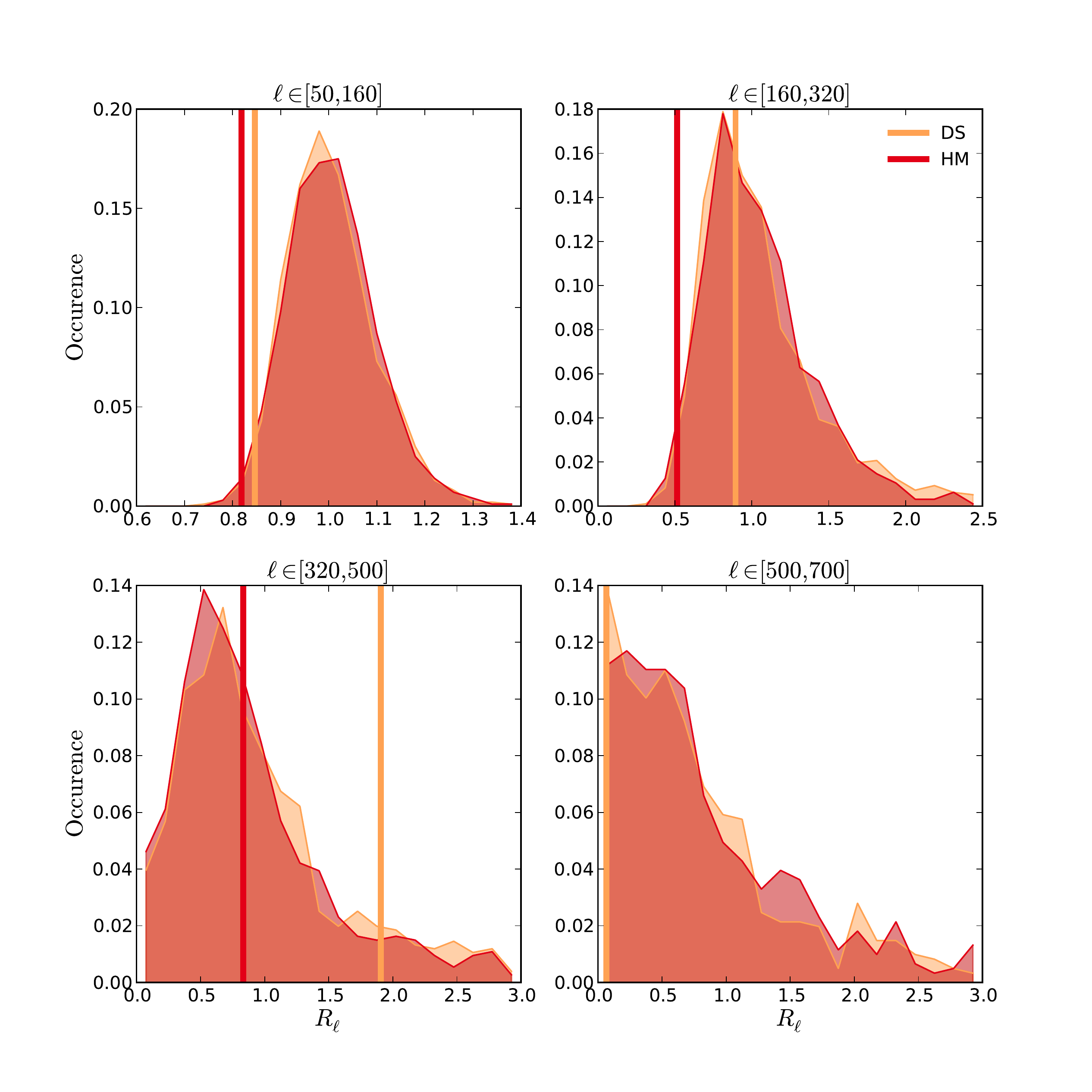} 
\caption{Same as Fig.~\ref{fig:distribs02}, for the LR24 region.}
\label{fig:distribs03}
\end{figure}

\begin{figure}
\includegraphics[width=0.5\textwidth]{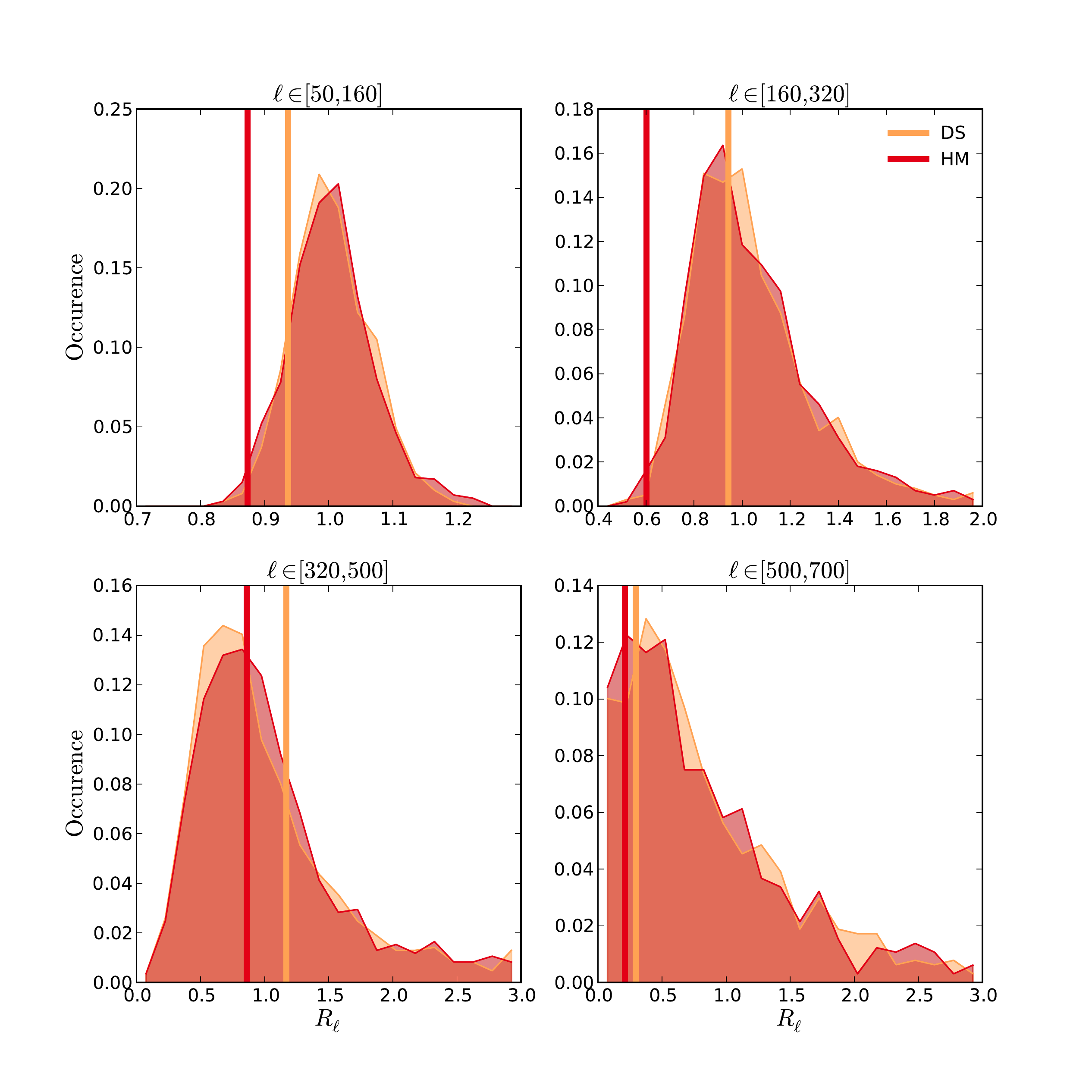} 
\caption{Same as Fig.~\ref{fig:distribs02}, for the LR33 region.}
\label{fig:distribs04}
\end{figure}

\begin{figure}
\includegraphics[width=0.5\textwidth]{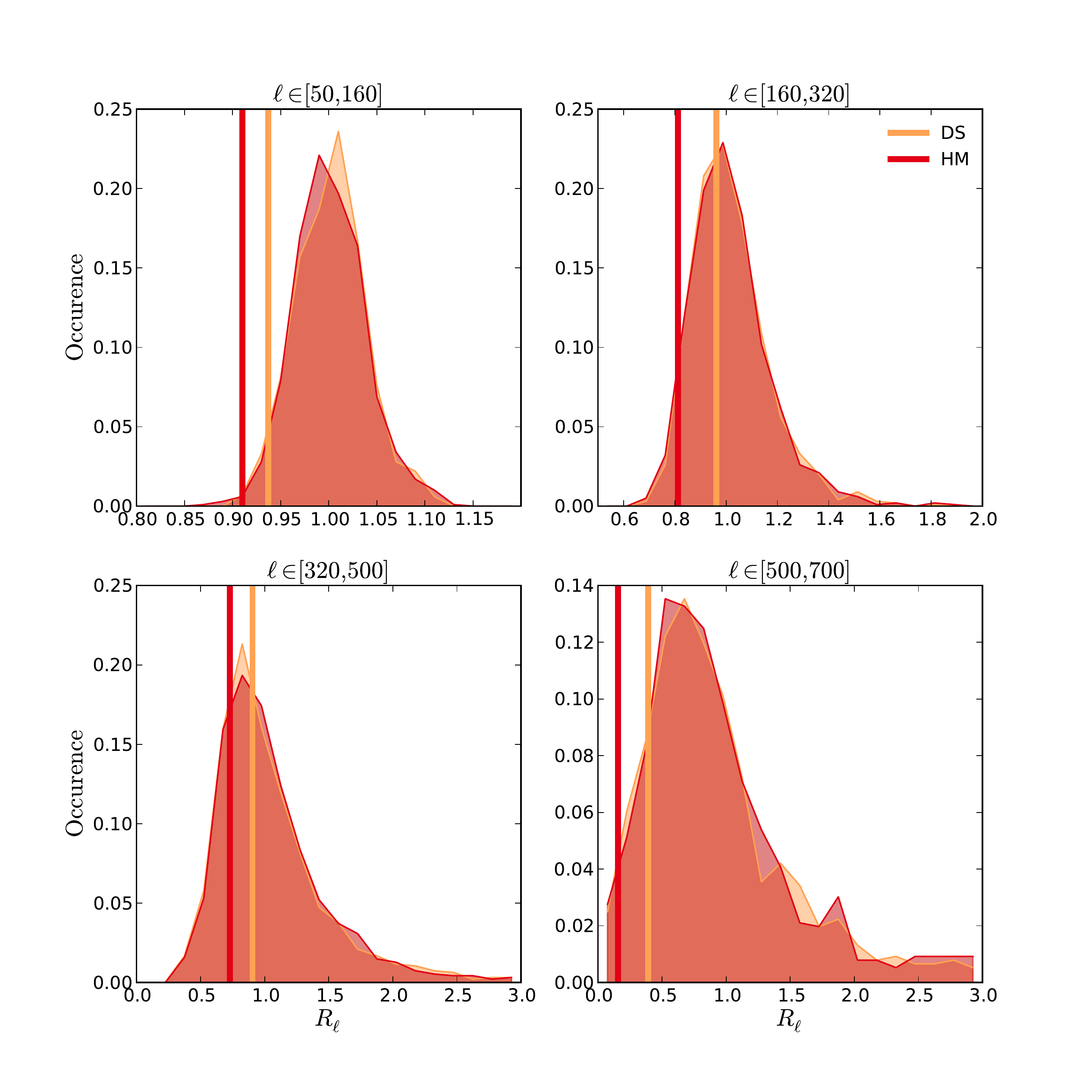} 
\caption{Same as Fig.~\ref{fig:distribs02}, for the LR42 region.}
\label{fig:distribs05}
\end{figure}

\begin{figure}
\includegraphics[width=0.5\textwidth]{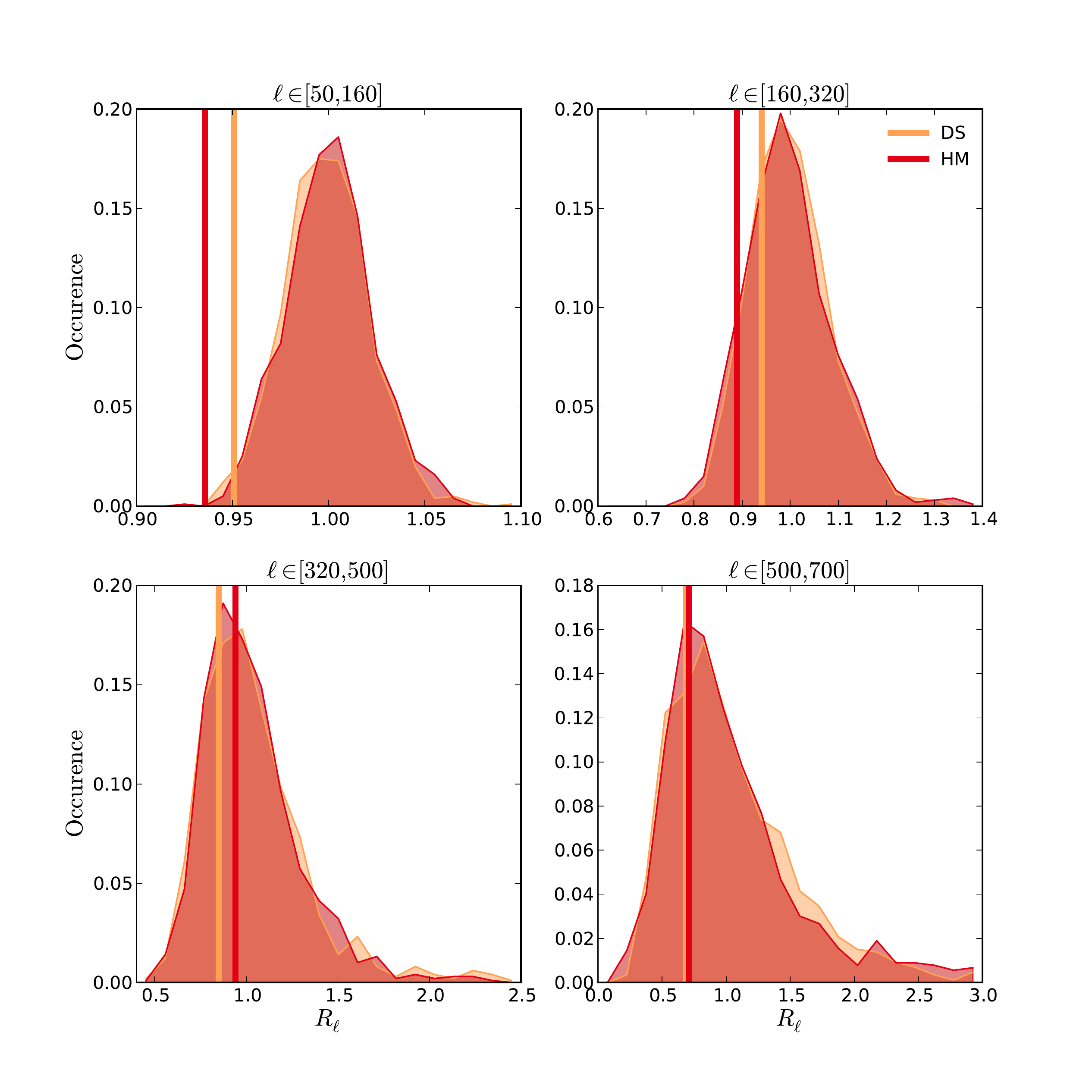} 
\caption{Same as Fig.~\ref{fig:distribs02}, for the LR53 region.}
\label{fig:distribs06}
\end{figure}

\begin{figure}
\includegraphics[width=0.5\textwidth]{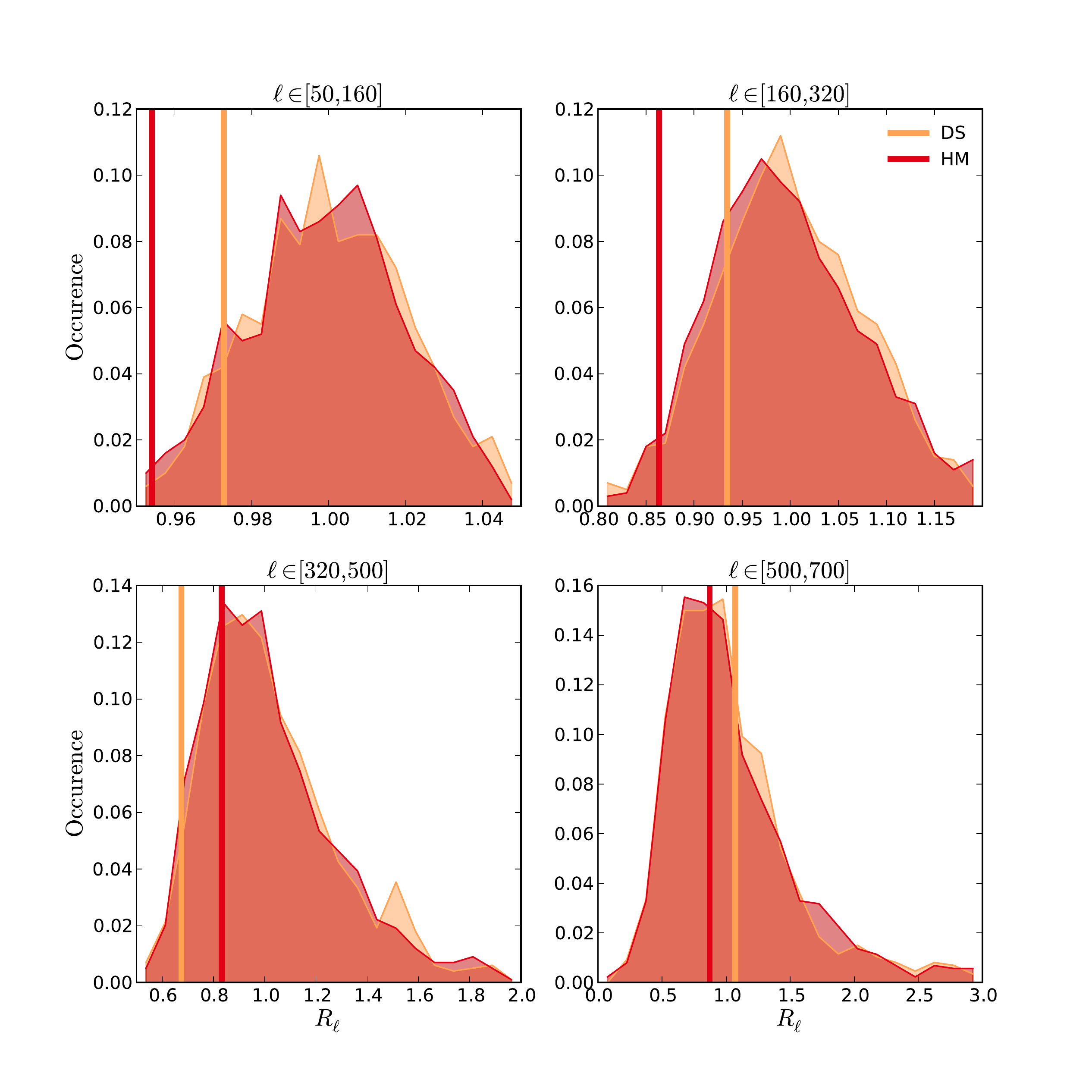} 
\caption{Same as Fig.~\ref{fig:distribs02}, for the LR63N region.}
\label{fig:distribs07N}
\end{figure}

\begin{figure}
\includegraphics[width=0.5\textwidth]{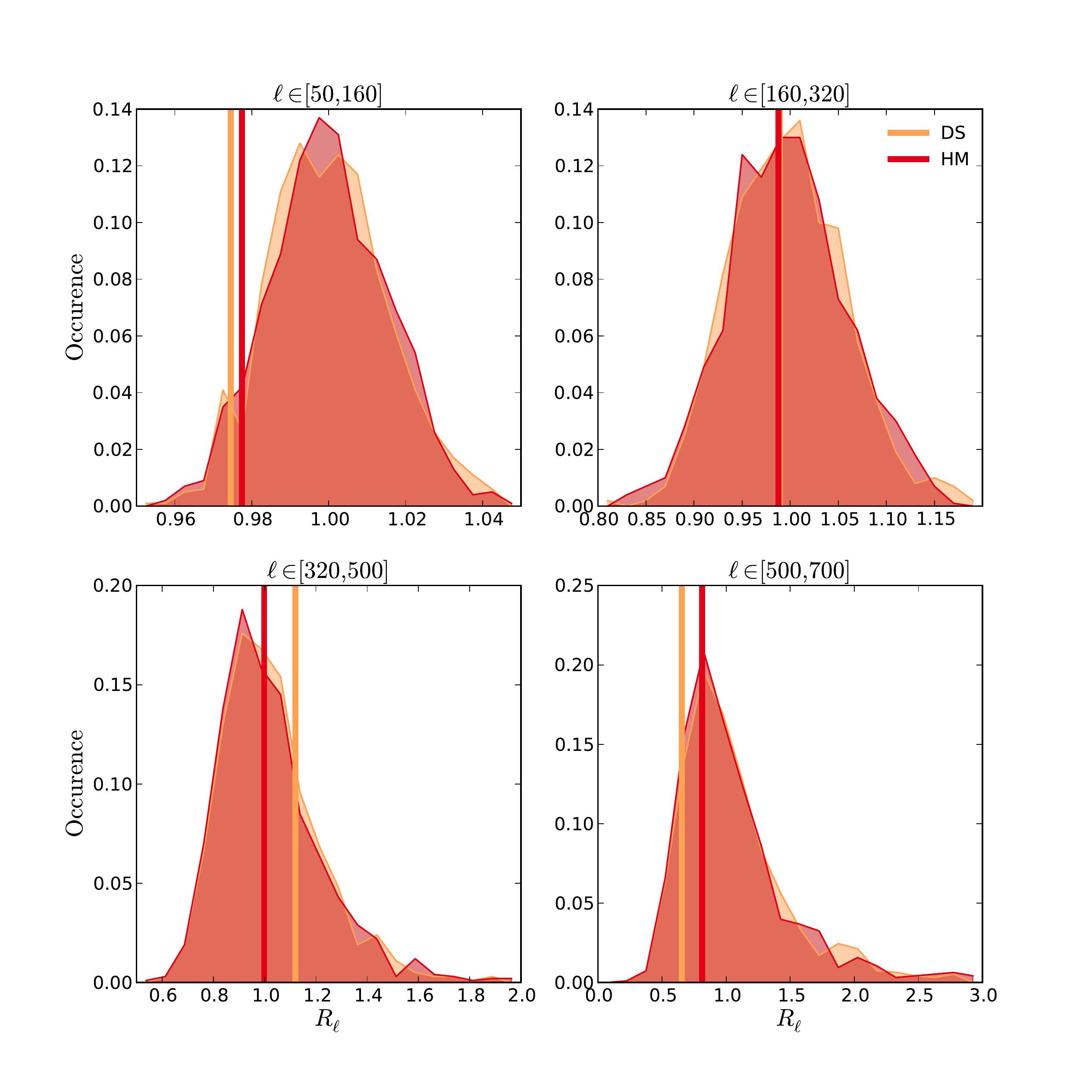} 
\caption{Same as Fig.~\ref{fig:distribs02}, for the LR63S region.}
\label{fig:distribs07S}
\end{figure}

\begin{figure}
\includegraphics[width=0.5\textwidth]{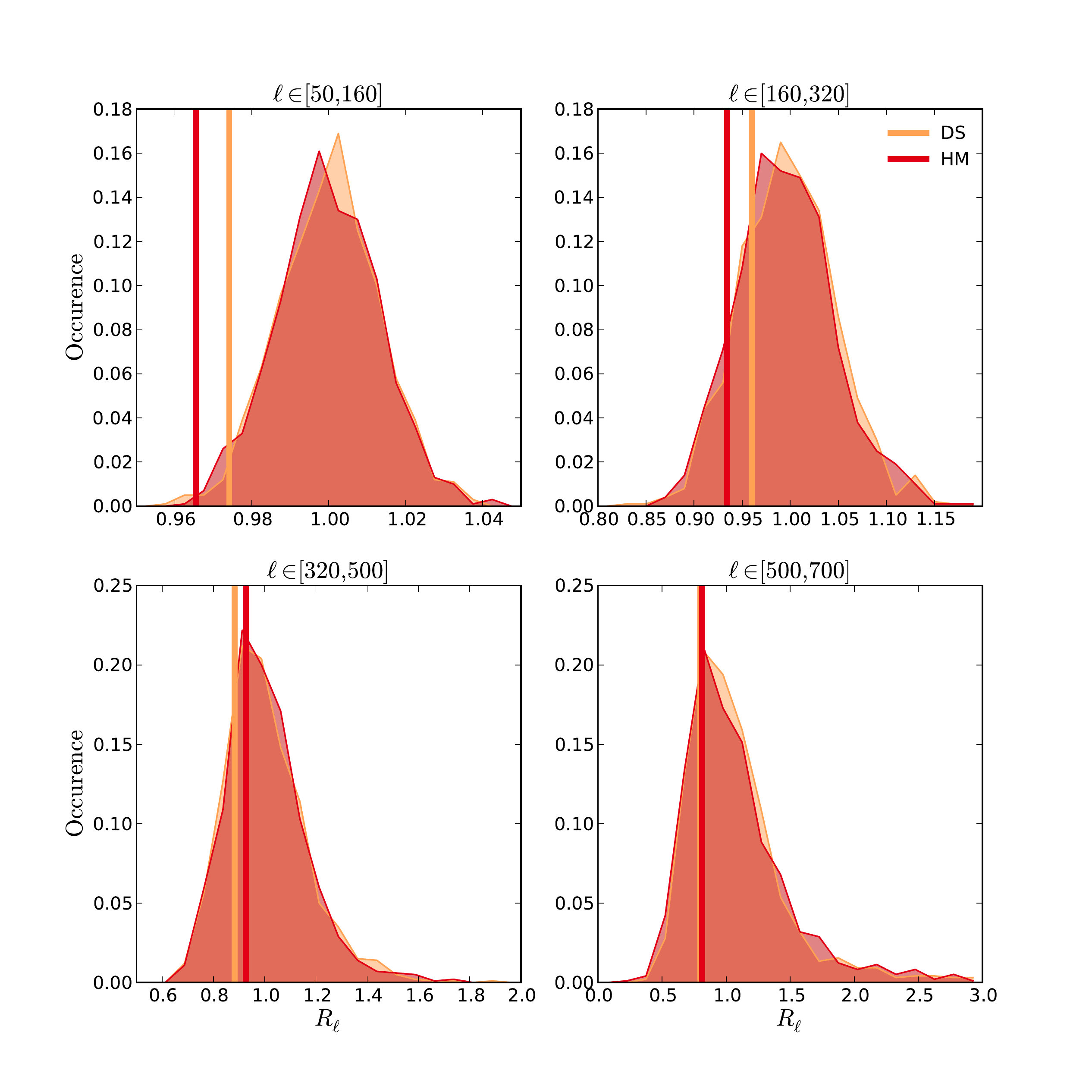} 
\caption{Same as Fig.~\ref{fig:distribs02}, for the LR63 region.}
\label{fig:distribs07}
\end{figure}

\begin{figure}
\includegraphics[width=0.5\textwidth]{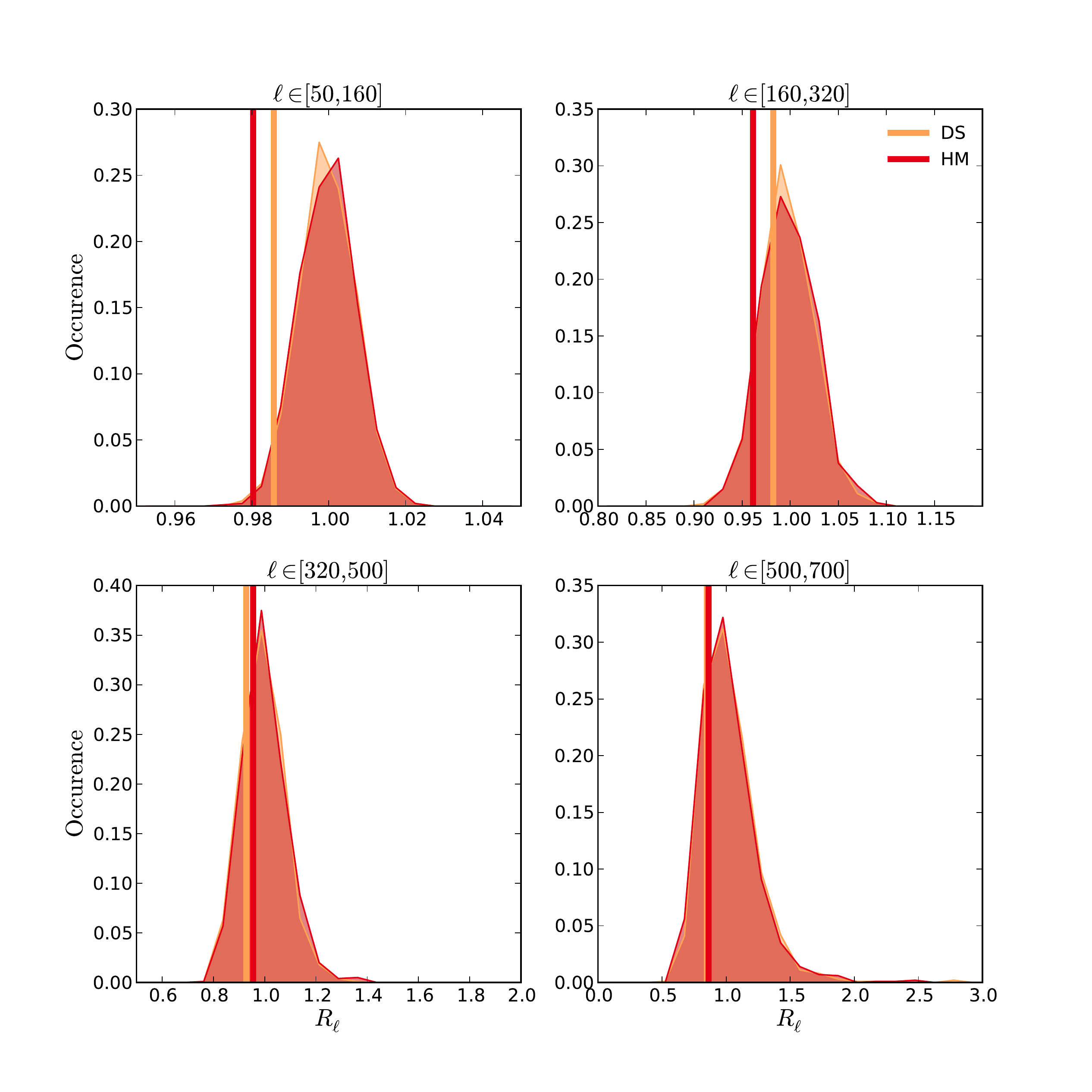} 
\caption{Same as Fig.~\ref{fig:distribs02}, for the LR72 region.}
\label{fig:distribs08}
\end{figure}

In Figs.~\ref{fig:distribs02}--\ref{fig:distribs08}, we show the histogram of the $\mathcal{R}_\ell^{BB}$ correlation ratio from 1\,000 realizations of detector-set and half-mission simulations. The histogram ``occurrences'' are the fractions of simulations that fall in each $\mathcal{R}_\ell^{BB}$ bin. These are compared to the values obtained from the \Planck\ data. Since we compute only noise-independent cross-spectra, when $\mathcal{R}_\ell^{BB}$ has a negative denominator, the data value is absent. The probability to exceed (PTE) for each data set, computed as the percentage of simulations that have a smaller $\mathcal{R}_\ell^{BB}$ than the data, is reported in Table~\ref{tab:ptes}.

\end{document}